\documentclass[12pt,a4paper]{article}

\usepackage{ifthen} 
\newboolean{pdflatex}
\setboolean{pdflatex}{true} 

\newboolean{articletitles}
\setboolean{articletitles}{true} 

\newboolean{uprightparticles}
\setboolean{uprightparticles}{true} 

\usepackage{microtype}
\usepackage{lineno}  
\usepackage{xspace} 

\usepackage{graphicx}  
\usepackage{color}
\usepackage{colortbl}
\graphicspath{{./figs/}} 

\usepackage{graphpap}
\usepackage{rotating}

\usepackage{multirow}
\usepackage{verbatim}

\usepackage{amsmath} 
\usepackage{amssymb}
\usepackage{amsfonts}
\usepackage{upgreek} 

\newcommand*\patchAmsMathEnvironmentForLineno[1]{%
\expandafter\let\csname old#1\expandafter\endcsname\csname #1\endcsname
\expandafter\let\csname oldend#1\expandafter\endcsname\csname
end#1\endcsname
 \renewenvironment{#1}%
   {\linenomath\csname old#1\endcsname}%
   {\csname oldend#1\endcsname\endlinenomath}%
}
\newcommand*\patchBothAmsMathEnvironmentsForLineno[1]{%
  \patchAmsMathEnvironmentForLineno{#1}%
  \patchAmsMathEnvironmentForLineno{#1*}%
}
\AtBeginDocument{%
\patchBothAmsMathEnvironmentsForLineno{equation}%
\patchBothAmsMathEnvironmentsForLineno{align}%
\patchBothAmsMathEnvironmentsForLineno{flalign}%
\patchBothAmsMathEnvironmentsForLineno{alignat}%
\patchBothAmsMathEnvironmentsForLineno{gather}%
\patchBothAmsMathEnvironmentsForLineno{multline}%
}

\usepackage{hyperref}    
\usepackage[all]{hypcap} 

\def\lhcb {\mbox{LHCb}\xspace}
\def\ux85 {\mbox{UX85}\xspace}

\def\babar  {\mbox{BaBar}\xspace}
\def\belle  {\mbox{Belle}\xspace}

\ifthenelse{\boolean{uprightparticles}}%
{
 
 \def\Pgamma      {\ensuremath{\upgamma}\xspace}

 \def\Pvarepsilon {\ensuremath{\upvarepsilon}\xspace}

 \def\Pmu         {\ensuremath{\upmu}\xspace}

 \def\Ppi         {\ensuremath{\uppi}\xspace}

 \def\Pphi        {\ensuremath{\upphi}\xspace}                 
                  
 \def\Pchi        {\ensuremath{\upchi}\xspace}                 
 \def\Ppsi        {\ensuremath{\uppsi}\xspace}

 \def\PDelta      {\ensuremath{\Delta}\xspace}                 
 \def\PXi      {\ensuremath{\Xi}\xspace}                 
 \def\PLambda      {\ensuremath{\Lambda}\xspace}                 
 \def\PSigma      {\ensuremath{\Sigma}\xspace}                 
 \def\POmega      {\ensuremath{\Omega}\xspace}                 
 \def\PUpsilon      {\ensuremath{\Upsilon}\xspace}                 
 
 \def\PB      {\ensuremath{\mathrm{B}}\xspace}                 
                  
 \def\PD      {\ensuremath{\mathrm{D}}\xspace}

 \def\PJ      {\ensuremath{\mathrm{J}}\xspace}                 
 \def\PK      {\ensuremath{\mathrm{K}}\xspace}

 \def\PS      {\ensuremath{\mathrm{S}}\xspace}

 \def\PW      {\ensuremath{\mathrm{W}}\xspace}                 
 \def\PX      {\ensuremath{\mathrm{X}}\xspace}

 \def\Pa      {\ensuremath{\mathrm{a}}\xspace}                 
 \def\Pb      {\ensuremath{\mathrm{b}}\xspace}                 
 \def\Pc      {\ensuremath{\mathrm{c}}\xspace}                 
 \def\Pd      {\ensuremath{\mathrm{d}}\xspace}                 
 \def\Pe      {\ensuremath{\mathrm{e}}\xspace}                 
 \def\Pf      {\ensuremath{\mathrm{f}}\xspace}

 \def\Pi      {\ensuremath{\mathrm{i}}\xspace}

 \def\Pp      {\ensuremath{\mathrm{p}}\xspace}

 \def\Ps      {\ensuremath{\mathrm{s}}\xspace}

}
{
 
 \def\Pgamma      {\ensuremath{\gamma}\xspace}

 \def\Pvarepsilon {\ensuremath{\varepsilon}\xspace}

 \def\Pmu         {\ensuremath{\mu}\xspace}

 \def\Ppi         {\ensuremath{\pi}\xspace}

 \def\Pphi        {\ensuremath{\phi}\xspace}                 
                  
 \def\Pchi        {\ensuremath{\chi}\xspace}                 
 \def\Ppsi        {\ensuremath{\psi}\xspace}                 
                  
 \mathchardef\PDelta="7101
 \mathchardef\PXi="7104
 \mathchardef\PLambda="7103
 \mathchardef\PSigma="7106
 \mathchardef\POmega="710A
 \mathchardef\PUpsilon="7107
                  
 \def\PB      {\ensuremath{B}\xspace}                 
                  
 \def\PD      {\ensuremath{D}\xspace}

 \def\PJ      {\ensuremath{J}\xspace}                 
 \def\PK      {\ensuremath{K}\xspace}

 \def\PS      {\ensuremath{S}\xspace}

 \def\PW      {\ensuremath{W}\xspace}                 
 \def\PX      {\ensuremath{X}\xspace}

 \def\Pa      {\ensuremath{a}\xspace}                 
 \def\Pb      {\ensuremath{b}\xspace}                 
 \def\Pc      {\ensuremath{c}\xspace}                 
 \def\Pd      {\ensuremath{d}\xspace}                 
 \def\Pe      {\ensuremath{e}\xspace}                 
 \def\Pf      {\ensuremath{f}\xspace}

 \def\Pi      {\ensuremath{i}\xspace}

 \def\Pp      {\ensuremath{p}\xspace}

 \def\Ps      {\ensuremath{s}\xspace}

}

\def\mumu       {\ensuremath{\Pmu^+\Pmu^-}\xspace}

\def\g      {\ensuremath{\Pgamma}\xspace}

\def\Wp     {\ensuremath{\PW^+}\xspace}

\def\dquark    {\ensuremath{\Pd}\xspace}

\def\squark    {\ensuremath{\Ps}\xspace}
\def\squarkbar {\ensuremath{\overline \squark}\xspace}

\def\cquark    {\ensuremath{\Pc}\xspace}
\def\cquarkbar {\ensuremath{\overline \cquark}\xspace}

\def\bquark    {\ensuremath{\Pb}\xspace}
\def\bquarkbar {\ensuremath{\overline \bquark}\xspace}

\def\pion  {\ensuremath{\Ppi}\xspace}
\def\piz   {\ensuremath{\pion^0}\xspace}

\def\pim   {\ensuremath{\pion^-}\xspace}

\def\kaon  {\ensuremath{\PK}\xspace}
  \def\Kbar  {\kern 0.2em\overline{\kern -0.2em \PK}{}\xspace}

\def\Kz    {\ensuremath{\kaon^0}\xspace}
\def\Kzb   {\ensuremath{\Kbar^0}\xspace}
\def\KzKzb {\ensuremath{\Kz \kern -0.16em \Kzb}\xspace}
\def\Kp    {\ensuremath{\kaon^+}\xspace}
\def\Km    {\ensuremath{\kaon^-}\xspace}

\def\KpKm  {\ensuremath{\Kp \kern -0.16em \Km}\xspace}

\def\Kstarz  {\ensuremath{\kaon^{*0}}\xspace}

\def\Kstarp  {\ensuremath{\kaon^{*+}}\xspace}

  \def\Dbar    {\kern 0.2em\overline{\kern -0.2em \PD}{}\xspace}
\def\D       {\ensuremath{\PD}\xspace}

\def\Dz      {\ensuremath{\D^0}\xspace}
\def\Dzb     {\ensuremath{\Dbar^0}\xspace}
\def\DzDzb   {\ensuremath{\Dz {\kern -0.16em \Dzb}}\xspace}
\def\Dp      {\ensuremath{\D^+}\xspace}
\def\Dm      {\ensuremath{\D^-}\xspace}

\def\DpDm    {\ensuremath{\Dp {\kern -0.16em \Dm}}\xspace}

\def\B       {\ensuremath{\PB}\xspace}
\def\Bbar    {\ensuremath{\kern 0.18em\overline{\kern -0.18em \PB}{}}\xspace}

\def\Bu      {\ensuremath{\B^+}\xspace}

\def\Bp      {\ensuremath{\Bu}\xspace}

\def\Bd      {\ensuremath{\B^0}\xspace}
\def\Bs      {\ensuremath{\B^0_\squark}\xspace}

\def\jpsi     {\ensuremath{{\PJ\mskip -3mu/\mskip -2mu\Ppsi\mskip 2mu}}\xspace}

\def\chiczero {\ensuremath{\Pchi_{\cquark 0}}\xspace}
\def\chicone  {\ensuremath{\Pchi_{\cquark 1}}\xspace}
\def\chictwo  {\ensuremath{\Pchi_{\cquark 2}}\xspace}
  \def\Y#1S{\ensuremath{\PUpsilon{(#1S)}}\xspace}

\def\chic  {\ensuremath{\Pchi_{c}}\xspace}

\def\proton      {\ensuremath{\Pp}\xspace}

\def\Lbar {\ensuremath{\kern 0.1em\overline{\kern -0.1em\PLambda}}\xspace}

\def\BF         {{\ensuremath{\cal B}\xspace}}

\def\BR         {\BF}

\def\to                 {\ensuremath{\rightarrow}\xspace}

\def\CP                {\ensuremath{C\!P}\xspace}

\def\AT#1     {\ensuremath{A_{\mathrm{T}}^{#1}}\xspace}           

\def\C#1      {\ensuremath{\mathcal{C}_{#1}}\xspace}                       
\def\Cp#1     {\ensuremath{\mathcal{C}_{#1}^{'}}\xspace}                    
\def\Ceff#1   {\ensuremath{\mathcal{C}_{#1}^{\mathrm{(eff)}}}\xspace}        
\def\Cpeff#1  {\ensuremath{\mathcal{C}_{#1}^{'\mathrm{(eff)}}}\xspace}       
\def\Ope#1    {\ensuremath{\mathcal{O}_{#1}}\xspace}                       
\def\Opep#1   {\ensuremath{\mathcal{O}_{#1}^{'}}\xspace}                    

\newcommand{\tev}{\ensuremath{\mathrm{\,Te\kern -0.1em V}}\xspace}
\newcommand{\gev}{\ensuremath{\mathrm{\,Ge\kern -0.1em V}}\xspace}
\newcommand{\mev}{\ensuremath{\mathrm{\,Me\kern -0.1em V}}\xspace}
\newcommand{\kev}{\ensuremath{\mathrm{\,ke\kern -0.1em V}}\xspace}
\newcommand{\ev}{\ensuremath{\mathrm{\,e\kern -0.1em V}}\xspace}
\newcommand{\gevc}{\ensuremath{{\mathrm{\,Ge\kern -0.1em V\!/}c}}\xspace}
\newcommand{\mevc}{\ensuremath{{\mathrm{\,Me\kern -0.1em V\!/}c}}\xspace}
\newcommand{\gevcc}{\ensuremath{{\mathrm{\,Ge\kern -0.1em V\!/}c^2}}\xspace}
\newcommand{\gevgevcccc}{\ensuremath{{\mathrm{\,Ge\kern -0.1em V^2\!/}c^4}}\xspace}
\newcommand{\mevcc}{\ensuremath{{\mathrm{\,Me\kern -0.1em V\!/}c^2}}\xspace}

\def\mum  {\ensuremath{\,\upmu\rm m}\xspace}

\def\invfb   {\ensuremath{\mbox{\,fb}^{-1}}\xspace}

\newcommand{\stat}{\ensuremath{\mathrm{(stat)}}\xspace}
\newcommand{\syst}{\ensuremath{\mathrm{(syst)}}\xspace}

\newcommand{\chisq}{\ensuremath{\chi^2}\xspace}

\def\gsim{{~\raise.15em\hbox{$>$}\kern-.85em
          \lower.35em\hbox{$\sim$}~}\xspace}
\def\lsim{{~\raise.15em\hbox{$<$}\kern-.85em
          \lower.35em\hbox{$\sim$}~}\xspace}

\def\pt         {\mbox{$p_{\rm T}$}\xspace}

\newcommand{\lum} {\ensuremath{\mathcal{L}}\xspace}

\def\evtgen     {\mbox{\textsc{EvtGen}}\xspace}
\def\pythia     {\mbox{\textsc{Pythia}}\xspace}

\def\geant      {\mbox{\textsc{Geant4}}\xspace}

\def\photos     {\mbox{\textsc{Photos}}\xspace}

\def\tell1  {TELL1\xspace}
\def\ukl1   {UKL1\xspace}

\usepackage{cite} 
\usepackage{mciteplus}

\usepackage{longtable} 

\begin{document}

\renewcommand{\thefootnote}{\fnsymbol{footnote}}
\setcounter{footnote}{1}


\begin{titlepage}
\pagenumbering{roman}

\vspace*{-1.5cm}
\centerline{\large EUROPEAN ORGANIZATION FOR NUCLEAR RESEARCH (CERN)}
\vspace*{1.5cm}
\hspace*{-0.5cm}
\begin{tabular*}{\linewidth}{lc@{\extracolsep{\fill}}r}
\ifthenelse{\boolean{pdflatex}}
{\vspace*{-2.7cm}\mbox{\!\!\!\includegraphics[width=.14\textwidth]{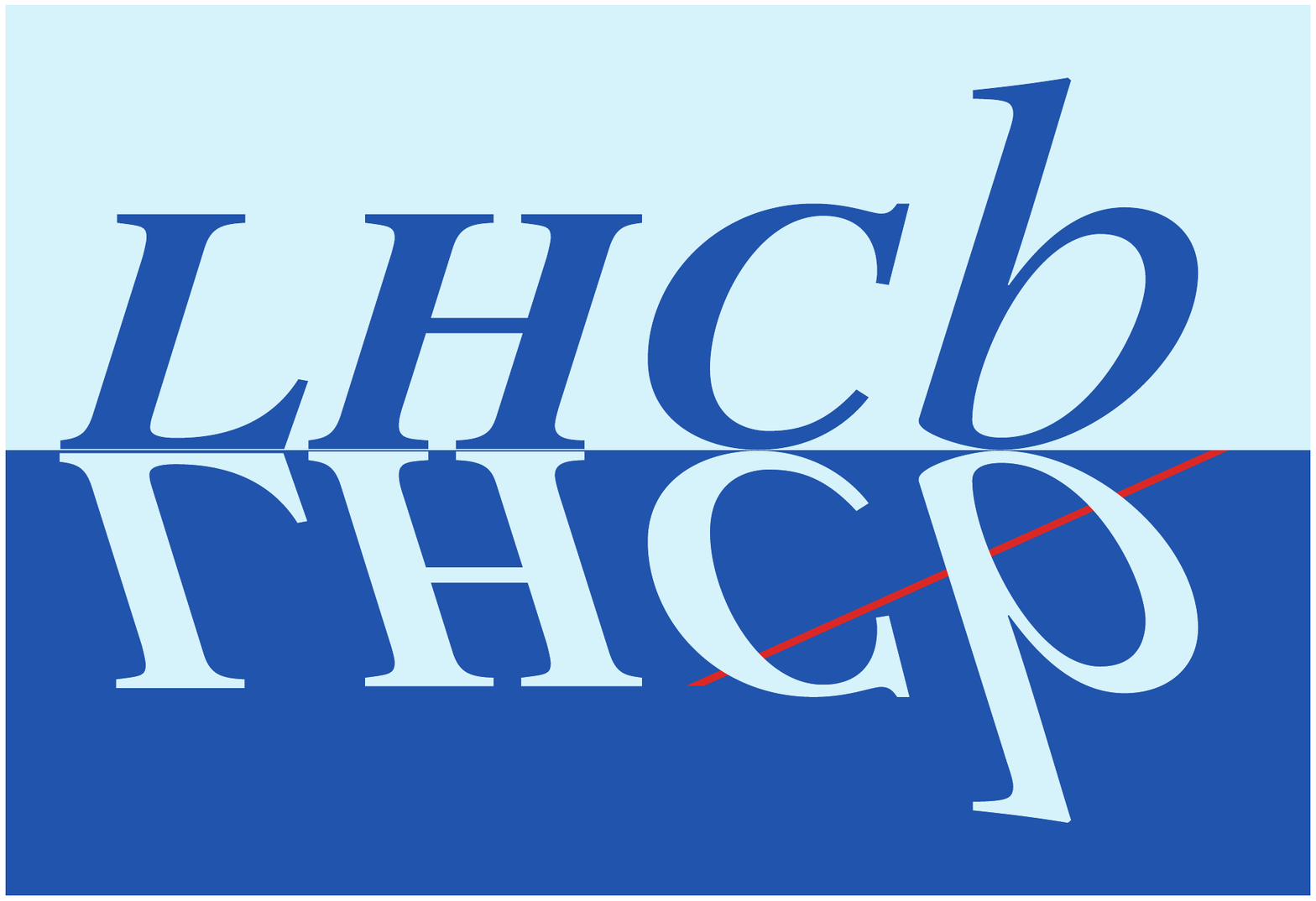}} & &}%
{\vspace*{-1.2cm}\mbox{\!\!\!\includegraphics[width=.12\textwidth]{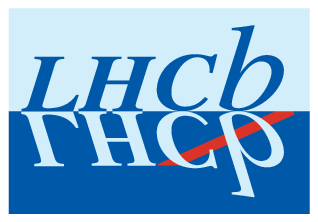}} & &}%
\\
 & & CERN-PH-EP-2013-088 \\  
 & & LHCb-PAPER-2013-024 \\  
 & & \today \\ 
\end{tabular*}

\vspace*{2.0cm}

{\bf\boldmath\huge
\begin{center}
  Observation of $\boldmath{\Bs\to\chicone\Pphi}$ decay and study of $\boldmath{\Bd\to\Pchi_{\Pc1,2}\Kstarz}$ decays
\end{center}
}

\vspace*{0.5cm}

\begin{center}
The LHCb collaboration\footnote{Authors are listed on the following pages.}
\end{center}

\vspace{0.5cm}

\begin{abstract}

\noindent The first observation of the decay ${\Bs\to\Pchi_{\Pc1}\Pphi}$ and a study of ${\Bd\to\Pchi_{\Pc1,2}\Kstarz}$ decays are presented. The 
analysis is performed using a dataset, corresponding to an integrated luminosity of 1.0 \invfb, collected by the \lhcb experiment in $\proton\proton$ 
collisions at a centre-of-mass energy of 7\tev. The following ratios of branching fractions are measured:


\begin{equation*}
\begin{array}{lll}

\dfrac{\BR(\Bs\to\Pchi_{\Pc1}\Pphi)}{\BR(\Bs\to\jpsi\Pphi)} &=& (18.9~\pm1.8\,\stat\pm1.3\,\syst\pm0.8\,(\BR)) \times 10^{-2},
\nonumber \\
\noalign{\vskip 3pt}
\dfrac{\BR(\Bd\to\Pchi_{\Pc1}\Kstarz)}{\BR(\Bd\to\jpsi\Kstarz)} &=& (19.8~\pm1.1\,\stat\pm1.2\,\syst\pm0.9\,(\BR)) \times 10^{-2},
\nonumber
\\
\noalign{\vskip 3pt}
\dfrac{\BR(\Bd\to\Pchi_{\Pc2}\Kstarz)}{\BR(\Bd\to\Pchi_{\Pc1}\Kstarz)} &=& (17.1~\pm5.0\,\stat\pm1.7\,\syst\pm1.1\,(\BR)) \times 10^{-2},
\nonumber
\\
\noalign{\vskip 3pt} 

\end{array}
\end{equation*} 


\noindent where the third uncertainty is due to the limited knowledge of the branching 
fractions of ${\Pchi_{\Pc}\to\jpsi\g}$ modes.

\end{abstract}

\vspace*{2.0cm}

\begin{center}

  Submitted to Nucl.~Phys.~B
\end{center}

\vspace{\fill}

{\footnotesize 
\centerline{\copyright~CERN on behalf of the \lhcb collaboration, license \href{http://creativecommons.org/licenses/by/3.0/}{CC-BY-3.0}.}}
\vspace*{2mm}

\end{titlepage}


\newpage
\setcounter{page}{2}
\mbox{~}
\newpage

\centerline{\large\bf LHCb collaboration}
\begin{flushleft}
\small
R.~Aaij$^{40}$, 
B.~Adeva$^{36}$, 
M.~Adinolfi$^{45}$, 
C.~Adrover$^{6}$, 
A.~Affolder$^{51}$, 
Z.~Ajaltouni$^{5}$, 
J.~Albrecht$^{9}$, 
F.~Alessio$^{37}$, 
M.~Alexander$^{50}$, 
S.~Ali$^{40}$
% ...

{\footnotesize \it
% ...
$ ^{3}$Center for High Energy Physics, Tsinghua University, Beijing, China\\
% ...
$ ^{5}$Clermont Universit\'{e}, Universit\'{e} Blaise Pascal, CNRS/IN2P3, LPC, Clermont-Ferrand, France\\
$ ^{6}$CPPM, Aix-Marseille Universit\'{e}, CNRS/IN2P3, Marseille, France\\
% ...
$ ^{9}$Fakult\"{a}t Physik, Technische Universit\"{a}t Dortmund, Dortmund, Germany\\
% ...
$ ^{36}$Universidad de Santiago de Compostela, Santiago de Compostela, Spain\\
$ ^{37}$European Organization for Nuclear Research (CERN), Geneva, Switzerland\\
% ...
$ ^{40}$Nikhef National Institute for Subatomic Physics, Amsterdam, The Netherlands\\
% ...
$ ^{45}$H.H. Wills Physics Laboratory, University of Bristol, Bristol, United Kingdom\\
% ...
$ ^{50}$School of Physics and Astronomy, University of Glasgow, Glasgow, United Kingdom\\
$ ^{51}$Oliver Lodge Laboratory, University of Liverpool, Liverpool, United Kingdom\\
% ...
}
\end{flushleft}

\cleardoublepage


\renewcommand{\thefootnote}{\arabic{footnote}}
\setcounter{footnote}{0}



\pagestyle{plain} 
\setcounter{page}{1}
\pagenumbering{arabic}


%

\section{Introduction}
\label{sec:Introduction}

\begin{figure}[htb]
  \setlength{\unitlength}{1mm}
  \centering
  \begin{picture}(60,30)

    \put(0,0){
      \includegraphics*[width=60mm,
      ]{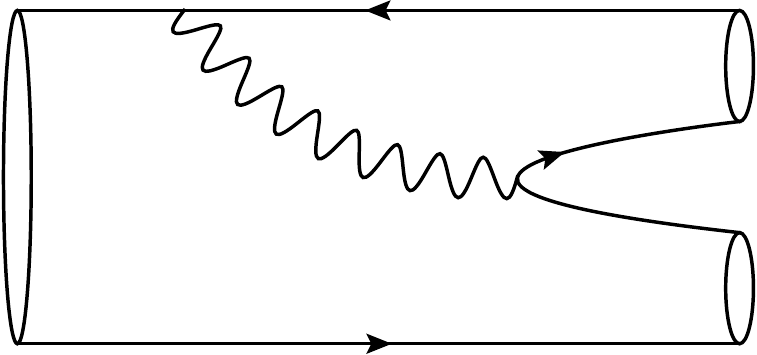}
    }
    \put(20,12){\Wp}

    \put(-19,13){$\Bd_{(\Ps)} \left\{ \begin{tabular}{r} 
\bquarkbar \, \\ \\ \\ \vspace{0.5em} \\
\dquark{(\squark)}  
\end{tabular} \right.$}

    \put(60,22){$\left.\begin{tabular}{l}
\cquarkbar ~~~\; \\
\cquark
\end{tabular}\right\} \Pchi_{\Pc}$}

    \put(60,4){$\left.\begin{tabular}{l}
\squarkbar \\
\dquark{(\squark)}
\end{tabular}\right\} \Kstarz(\Pphi)$}

  \end{picture}
  \caption { \small Leading-order tree level diagram for the $\Bd_{(\Ps)}\to\Pchi_{\Pc}\PX$ decays.}

  \label{fig:Diagram}
\end{figure}

Two-body \B-meson decays into a final states containing charmonium meson have played a crucial role in the observation of \CP violation in the 
\B-meson system. These decay 
modes also provide a sensitive laboratory for studying the effects of the strong interaction. Such decays 
are expected to proceed predominantly via the colour-suppressed tree diagram involving $\bquarkbar\to\cquarkbar\cquark\squarkbar$ transition shown 
in Fig.~\ref{fig:Diagram}. Under the factorization hypothesis the branching ratios of the ${\Bd_{(\Ps)}\to\Pchi_{\Pc0,2}\PX}$ decays, where $\PX$ 
denotes a \Kstarz or a $\Pphi$ meson, are expected to be small in comparison to $\Bd_{(\Ps)}\to\Pchi_{\Pc1}\PX$ decays~\cite{Puzzles}. 
However, non-factorizable contributions may be large~\cite{Puzzles}; the branching fraction for the $\Bd\to\chiczero\Kstarz$ decay was measured by the 
\babar collaboration to be $(1.7\pm 0.3\pm 0.2)\times 10^{-4}$~\cite{babarChic0Kstarz} while the branching fraction for the 
$\Bd\to\chicone\Kstarz$ decay was measured by the \babar and \belle collaborations to be 
$(2.5\pm0.2\pm0.2)\times 10^{-4}$~\cite{babar} and $(1.73^{+0.15 +0.34}_{-0.12 -0.22})\times 10^{-4}$~\cite{belle}, respectively.
The branching fraction for the decay $\Bd\to\chictwo\Kstarz$ has been measured by the \babar collaboration to be $(6.6\pm1.8\pm0.5)\times 
10^{-5}$~\cite{babar} and, unlike the branching fraction for the $\Bd\to\chiczero\Kstarz$ decay, can still be explained in the factorization 
approach~\cite{BenekeVernazza}. Therefore, future measurements of the branching fractions of both $\Bd\to\chicone\Kstarz$ and $\Bd\to\chictwo\Kstarz$ 
decays can provide valuable information for the understanding of the production of $\Pchi_{\Pc}$ states in \B meson decays, where $\Pchi_{\Pc}$ 
denotes \chicone and \chictwo states. The decay modes $\Bs\to\Pchi_{\Pc}\Pphi$ have not been observed previously. 

In this paper, the first observation of the decay $\Bs\to\chicone\Pphi$ and a study of the $\Bd\to\Pchi_{\Pc1,2}\Kstarz$ decays are presented. The 
analysis is based on a data sample, corresponding to an integrated luminosity of 1.0\invfb, collected with the \lhcb detector in 
$\proton\proton$ collisions at a centre-of-mass energy of 7\tev.

\section{LHCb detector}
\label{sec:Detector}

The \lhcb detector~\cite{Alves:2008zz} is a single-arm forward
spectrometer covering the \mbox{pseudorapidity} range $2<\eta <5$,
designed for the study of particles containing \bquark or \cquark quarks. The detector includes a high precision tracking system
consisting of a silicon-strip vertex detector surrounding the $\proton\proton$ 
interaction region, a large-area silicon-strip detector located
upstream of a dipole magnet with a bending power of about
$4{\rm\,Tm}$, and three stations of silicon-strip detectors and straw
drift tubes placed downstream. The combined tracking system has 
momentum resolution $\Delta p/p$ that varies from 0.4\% at 5\gevc to
0.6\% at 100\gevc, and impact parameter resolution of 20\mum for
tracks with high transverse momentum~(\pt). Charged hadrons are identified
using two ring-imaging Cherenkov detectors~\cite{arxiv:1211-6759}. Photon, electron and
hadron candidates are identified by a calorimeter system consisting of
scintillating-pad and preshower detectors, an electromagnetic
calorimeter and a hadronic calorimeter. Muons are identified by a
system composed of alternating layers of iron and multiwire
proportional chambers~\cite{Muon:performance}. 

The trigger~\cite{Aaij:2012me} consists of a 
hardware stage, based on information from the calorimeter and muon 
systems, followed by a software stage where a full event reconstruction is applied. 
Candidate events are first required to pass a hardware trigger 
which selects muons with $\pt>1.48\gevc$. In 
the subsequent software trigger, at least 
one of the muons is required to have both 
$\pt>0.8\gevc$ and impact parameter larger than $100\mum$ with respect to all 
of the primary $\proton\proton$ interaction vertices~(PVs) in the 
event. Finally, the two final state muons are required to form a vertex that is significantly 
displaced from the PVs.

The analysis technique reported below has been validated using simulated events. The $\proton\proton$ collisions are generated using 
\pythia~6.4~\cite{Sjostrand:2006za} with a specific \lhcb~configuration~\cite{LHCb-PROC-2010-056}.  
Decays of hadronic particles
are described by \evtgen~\cite{Lange:2001uf} in which final state
radiation is generated using \photos~\cite{Golonka:2005pn}. 
The interaction of the generated particles with the detector and its
response are implemented using the \geant toolkit~\cite{Agostinelli:2002hh,Allison:2006ve} 
as described in Ref.~\cite{LHCb-PROC-2011-006}.

\section{Event selection}
\label{sec:EventSelection}

The decays $\Bd\to\Pchi_{\Pc}\Kstarz$ and $\Bs\to\Pchi_{\Pc}\Pphi$ (the inclusion of charged conjugate processes is implied throughout) are 
reconstructed using the 
$\Pchi_{\Pc}\to\jpsi\g$ decay mode. The decays ${\Bd\to\jpsi\Kstarz}$ and $\Bs\to\jpsi\Pphi$ are used as normalization channels. The intermediate 
resonances are reconstructed in the ${\jpsi\to\mumu}$, ${\Kstarz\to\Kp\pim}$ and ${\Pphi\to\Kp\Km}$ final states. 

As in Refs.~\cite{B2psiX_paper,Dasha_paper,Serezha_paper}, 
pairs of oppositely-charged tracks identified as muons, each having $\pt>0.55\gevc$ and originating from a common vertex, are combined to form 
$\jpsi\to\mumu$ candidates. Track quality is ensured by requiring the \chisq per number of degrees of freedom (\chisq/ndf) provided by the track fit 
to be less than~5. Well identified muons are selected by requiring that the difference in logarithms of the likelihood of the muon hypothesis  
with respect to the hadron hypothesis is 
larger than zero~\cite{Muon:performance}. The fit of the common two-prong vertex is required to satisfy $\chisq/{\rm ndf}<20$. The vertex is required 
to be well separated 
from the reconstructed primary vertex of any of the $\proton\proton$ interactions by requiring the decay length to be at least three  
times its uncertainty. Finally, the invariant mass of the dimuon combination is required to be between 3.020 and~3.135~\gevcc.

To create $\Pchi_{\Pc}$ candidates, the selected $\jpsi$ candidates are combined with a photon that has been reconstructed using clusters in 
the electromagnetic calorimeter that have transverse energy greater than 0.7\gev. To suppress the large combinatorial background from 
$\piz\to\g\g$ decays, photons that can form part of a $\piz\to\g\g$ candidate with invariant mass within $10\mevcc$ of the known $\piz$ 
mass~\cite{PDG} are not used for reconstruction of $\Pchi_{\Pc}$ candidates. To be considered as a $\Pchi_{\Pc}$, the $\jpsi\g$ combination needs to 
have a transverse momentum larger than 3\gevc and an invariant mass in the range 3.4 -- 3.7\gevcc.

The selected $\Pchi_{\Pc}$ and $\jpsi$ candidates are then combined with $\Kp\pim$ or $\Kp\Km$ pairs to create $\Bd_{(\Ps)}$ meson candidates. 
To identify kaons (pions), the difference in logarithm of the likelihood of the kaon and pion hypotheses~\cite{arxiv:1211-6759} is required to be 
greater than (less than) zero. 
The track $\chisq/\rm{ndf}$ provided by the track fit is required to be less than 5. The kaons and pions are required to have transverse momentum 
larger than 0.8\gevc and to have an impact parameter \chisq
, defined as the difference between the \chisq of the reconstructed $\proton\proton$ collision vertex formed with and without the considered track, 
larger than~4. The invariant mass of the kaon and pion system, $M_{\Kp\pim}$, is required to be 
${0.675<M_{\Kp\pim}<1.215\gevcc}$ and the invariant mass of the kaon pair, $M_{\Kp\Km}$, is required to be ${0.999<M_{\Kp\Km}<1.051\gevcc}$. In 
the reconstruction of \Kstarz candidates, a possible background arises from $\Pphi\to\Kp\Km$ decays when a kaon is misidentified as a pion. To 
suppress this contribution, the invariant mass of the kaon and pion system, calculated under the kaon mass hypothesis for the pion track, is required 
to be outside the range from 1.01 to 1.03\gevcc.

In addition, the decay time of \B candidates is required to be larger than~150\mum/$c$ to reduce the large combinatorial background from 
particles produced in the primary $\proton\proton$ interaction. To improve the invariant mass resolution of the $\Bd_{(s)}$ meson candidate a 
kinematic fit~\cite{DTF} is performed. In this fit, constraints are applied to the masses of the intermediate \jpsi and $\Pchi_{\Pc}$ 
resonances~\cite{PDG} and it is also required that the $\Bd_{(\Ps)}$ meson candidate momentum vector points to the primary vertex. The \chisq/ndf for 
this fit is required to be less than~5.

\section[ $\text{\Bd\!\!\to{\ensuremath{\Pchi_{\Pc}}\xspace}\Kstarz}$ and $\text{\Bs\!\!\to{\ensuremath{\Pchi_{\Pc1}}\xspace}\Pphi}$ decays ] { 
$\boldmath{\text{\Bd\!\!\to{\ensuremath{\Pchi_{\Pc}}\xspace}\Kstarz}}$ and $\boldmath{\text{\Bs\!\!\to{\ensuremath{\Pchi_{\Pc1}}\xspace}\Pphi}}$ 
decays }

\begin{figure}[t]
  \setlength{\unitlength}{1mm}
  \centering

  \begin{picture}(160,80)  

    \put(0,40){
      \includegraphics*[width=80mm,
      ]{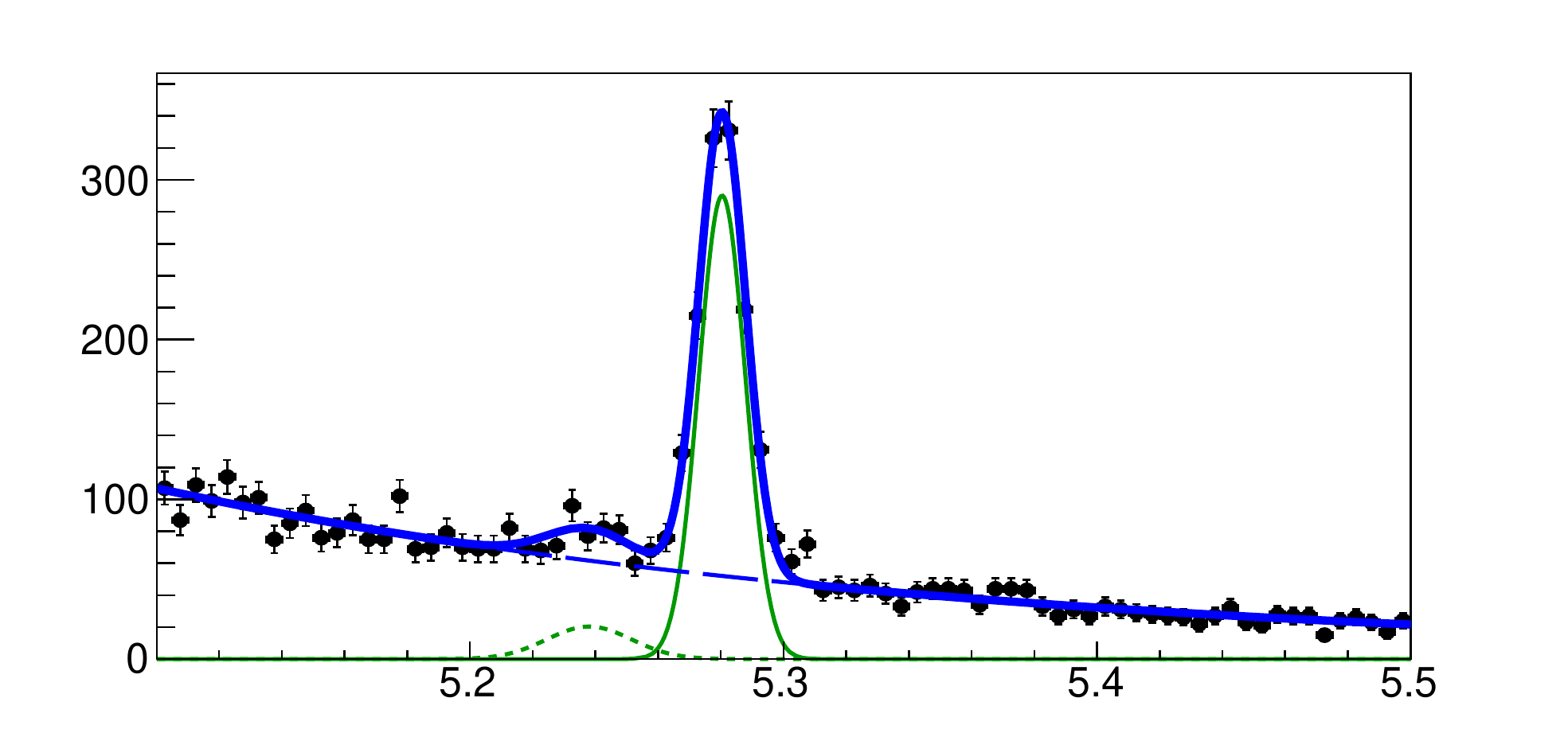}
    }
    \put(0,0){
      \includegraphics*[width=80mm,
      ]{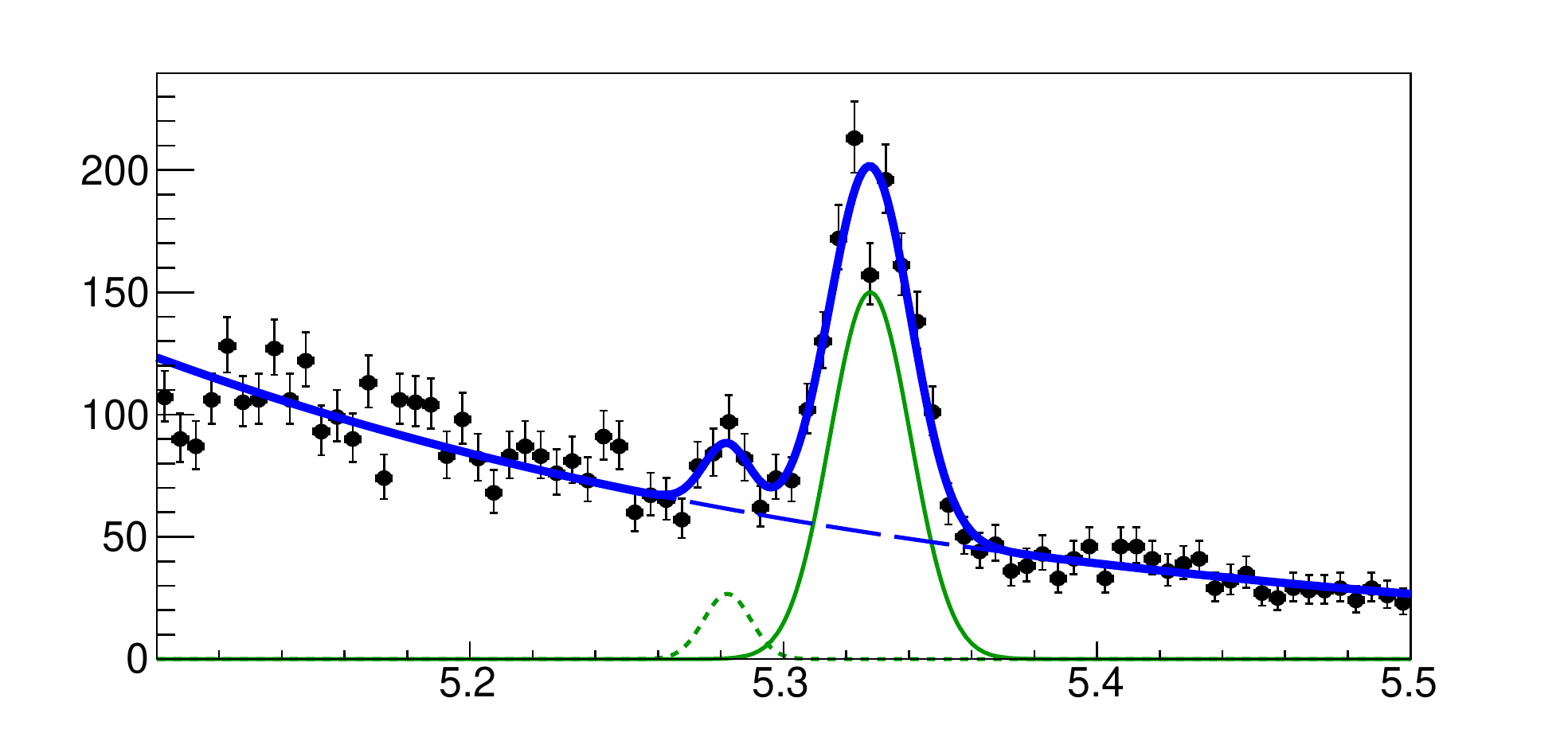}
    }
    \put(80,40){
      \includegraphics*[width=80mm,
      ]{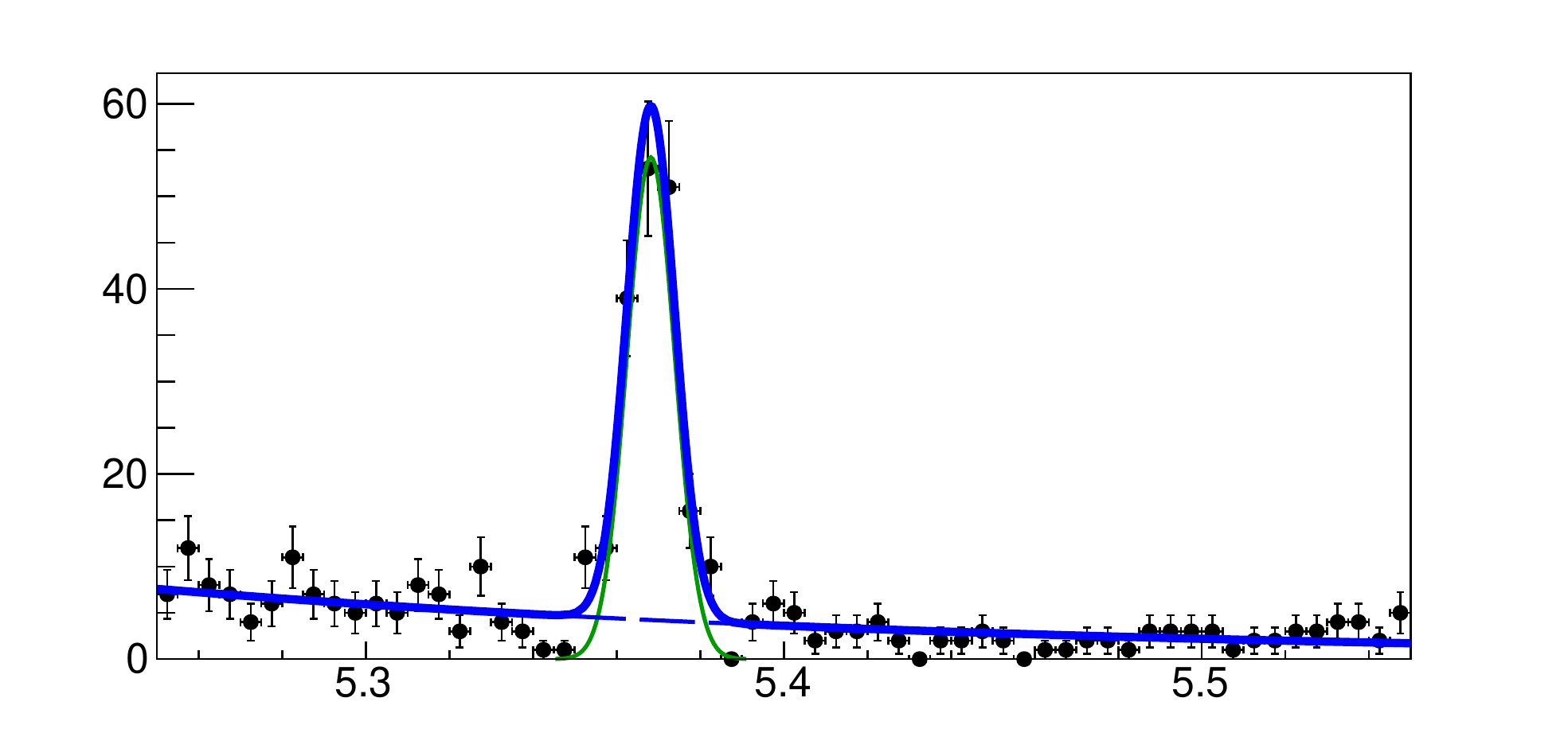}
    }
    \put(80,0){
      \includegraphics*[width=80mm,
      ]{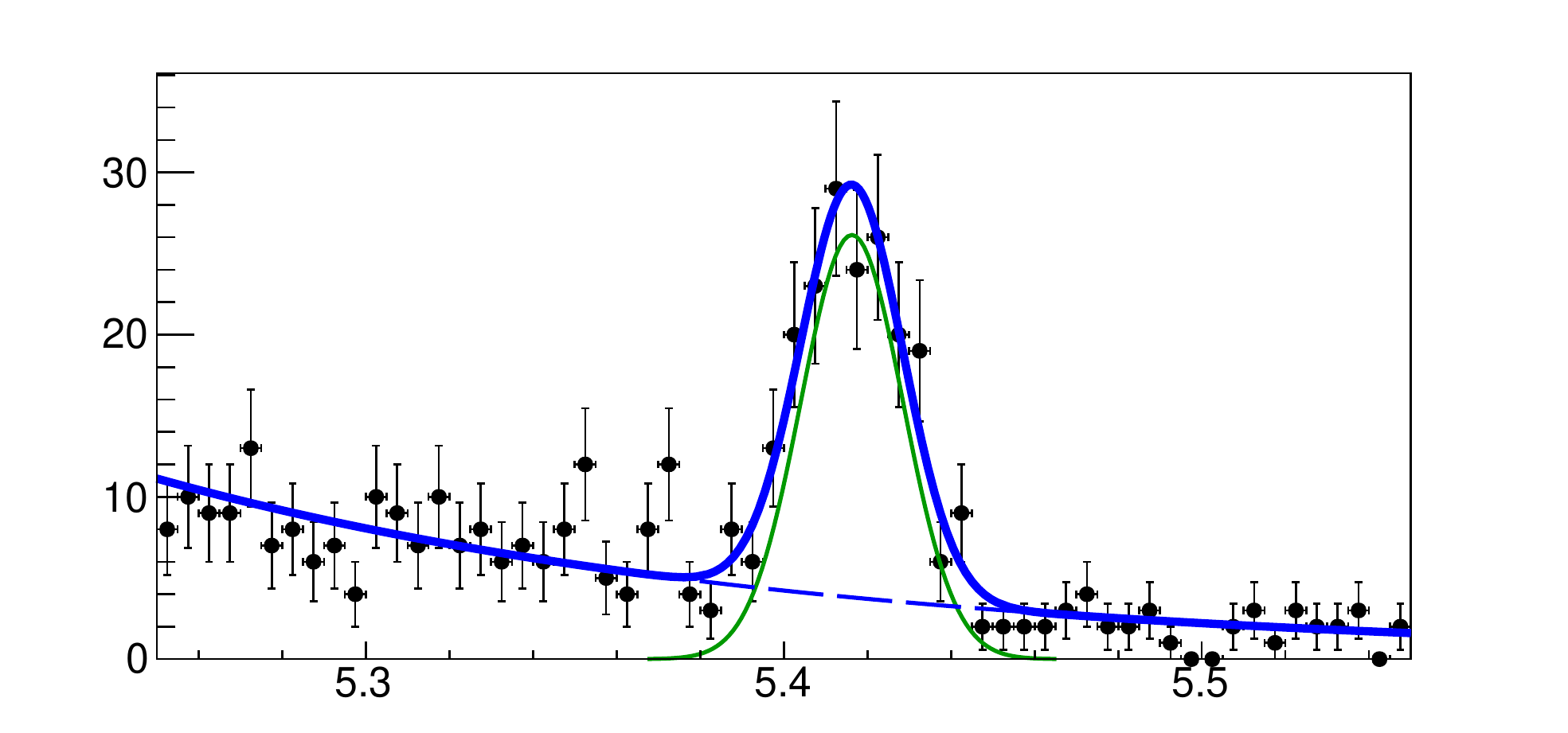}
    }

    \put(15,68){\small \lhcb}
    \put(95,68){\small \lhcb}
    \put(15,28){\small \lhcb}
    \put(95,28){\small \lhcb}

    \put(48,65){\small \footnotesize $\begin{array}{l} (\Pa) \\ \Bd\to\Pchi_{\Pc}\Kstarz \\ \rm{\chicone~constraint} \end{array}$}
    \put(128,65){\small \footnotesize $\begin{array}{l} (\Pb) \\ \Bs\to\chicone\Pphi \\ \rm{\chicone~constraint} \end{array}$}
    \put(48,25){\small \footnotesize $\begin{array}{l} (\Pc) \\ \Bd\to\Pchi_{\Pc}\Kstarz \\ \rm{\chictwo~constraint} \end{array}$}
    \put(128,25){\small \footnotesize $\begin{array}{l} (\Pd) \\ \Bs\to\chicone\Pphi \\ \rm{\chictwo~constraint} \end{array}$}

    \put(38,38){\scriptsize $\mathrm{M_{\Pchi_{\Pc}\Kstarz}}$}
    \put(38,-2){\scriptsize $\mathrm{M_{\Pchi_{\Pc}\Kstarz}}$}
    \put(118,38){\scriptsize $\mathrm{M_{\chicone\Pphi}}$}
    \put(118,-2){\scriptsize $\mathrm{M_{\chicone\Pphi}}$}

    \put(62,38){\scriptsize $\mathrm{[GeV/c^{2}]}$}
    \put(62,-2){\scriptsize $\mathrm{[GeV/c^{2}]}$}
    \put(142,38){\scriptsize $\mathrm{[GeV/c^{2}]}$}
    \put(142,-2){\scriptsize $\mathrm{[GeV/c^{2}]}$}

    \begin{sideways}

    \put(43,0){\scriptsize Candidates / (5\mevcc)}
    \put(43,80){\scriptsize Candidates / (5\mevcc)}
    \put(3,0){\scriptsize Candidates / (5\mevcc)}
    \put(3,80){\scriptsize Candidates / (5\mevcc)}

    \end{sideways}
    
  \end{picture}

  \caption { \small Invariant mass distributions for: (a) $\Bd\to\Pchi_{\Pc}\Kstarz$ and (b) $\Bs\to\chicone\Pphi$ candidates with $\chicone$ mass 
constraint; (c) $\Bd\to\Pchi_{\Pc}\Kstarz$ and (d) $\Bs\to\chicone\Pphi$ candidates with $\chictwo$ mass constraint. The total fitted function (thick 
solid blue), signal for the $\chicone$ and $\chictwo$ modes (thin green solid and dotted, respectively) and the combinatorial background (dashed blue) 
are shown.}

  \label{fig:MassDist} \end{figure}

The invariant mass distributions after selecting ${\Bd\to\Pchi_{\Pc}\Kstarz}$ and ${\Bs\to\chicone\Pphi}$ candidates, separately with a \chicone and 
\chictwo mass constraints, are shown in Fig.~\ref{fig:MassDist}. The signal is modelled by a single Gaussian function and the combinatorial background 
is modelled by an exponential function. In the \Bd channel (Figs.~2(a) and (c)), the right peak in the mass distributions corresponds to the \chicone 
mode and the left one to the \chictwo mode. Owing to the small $\chiczero\to\jpsi\g$ branching fraction~\cite{PDG} the contribution from the \chiczero 
mode is negligible. As the \Bd candidate mass is calculated with the $\jpsi\g$ invariant mass constrained to the $\chicone$ ($\chictwo$) known mass, 
the signal peak corresponding to the $\chictwo$ ($\chicone$) mode is shifted to a lower (higher) value with respect to the \Bd mass. The same effect 
is observed in simulation. The ratio of the mass resolutions of these two signal peaks is fixed to the value obtained from simulation. In the \Bs 
channel no significant contribution from the \chictwo decay mode is expected and therefore it is not considered in the fit. The statistical 
significance for the observed signal is determined as $S=\sqrt{-2\ln \frac{\lum_{\rm B}}{\lum_{\rm S+B}}}$, where $\lum_{\rm S+B}$ and $\lum_{\rm B}$ 
denote the likelihood of the signal plus background hypothesis and the background only hypothesis, respectively. The statistical significance of the 
$\Bs\to\chicone\Pphi$ signal is found to be larger than 9 standard deviations.

The positions and resolutions of the signal peaks are consistent with the expectations from simulation. To investigate the different signal yields 
obtained with the \chicone and \chictwo mass constraints, a simplified simulation study was performed, which accounts for correlations, differences in 
selection efficiencies and background fluctuations. This study demonstrates that the yields are in agreement within the statistical uncertainty.


To examine the resonance structure of the $\Bd\to\Pchi_{\Pc}\Kstarz$ and $\Bs\to\Pchi_{\Pc1}\Pphi$ decays, the {\em sPlot} technique~\cite{sPlot} was 
used with weights determined from the $\Bd_{(\Ps)}$ candidate invariant mass fits described above. The invariant mass 
distributions for each signal component are obtained. For the $\jpsi\g$ invariant mass distributions the requirement on the invariant mass of the 
$\Kp\pim$($\Kp\Km$) system is tightened to be within $50(10)\mevcc$ around the known $\Kstarz$($\Pphi$) mass to reduce background.

\begin{figure}[t]
  \setlength{\unitlength}{1mm}
  \centering
  \begin{picture}(160,145)

    \put(0,100){
      \includegraphics*[width=80mm,
      ]{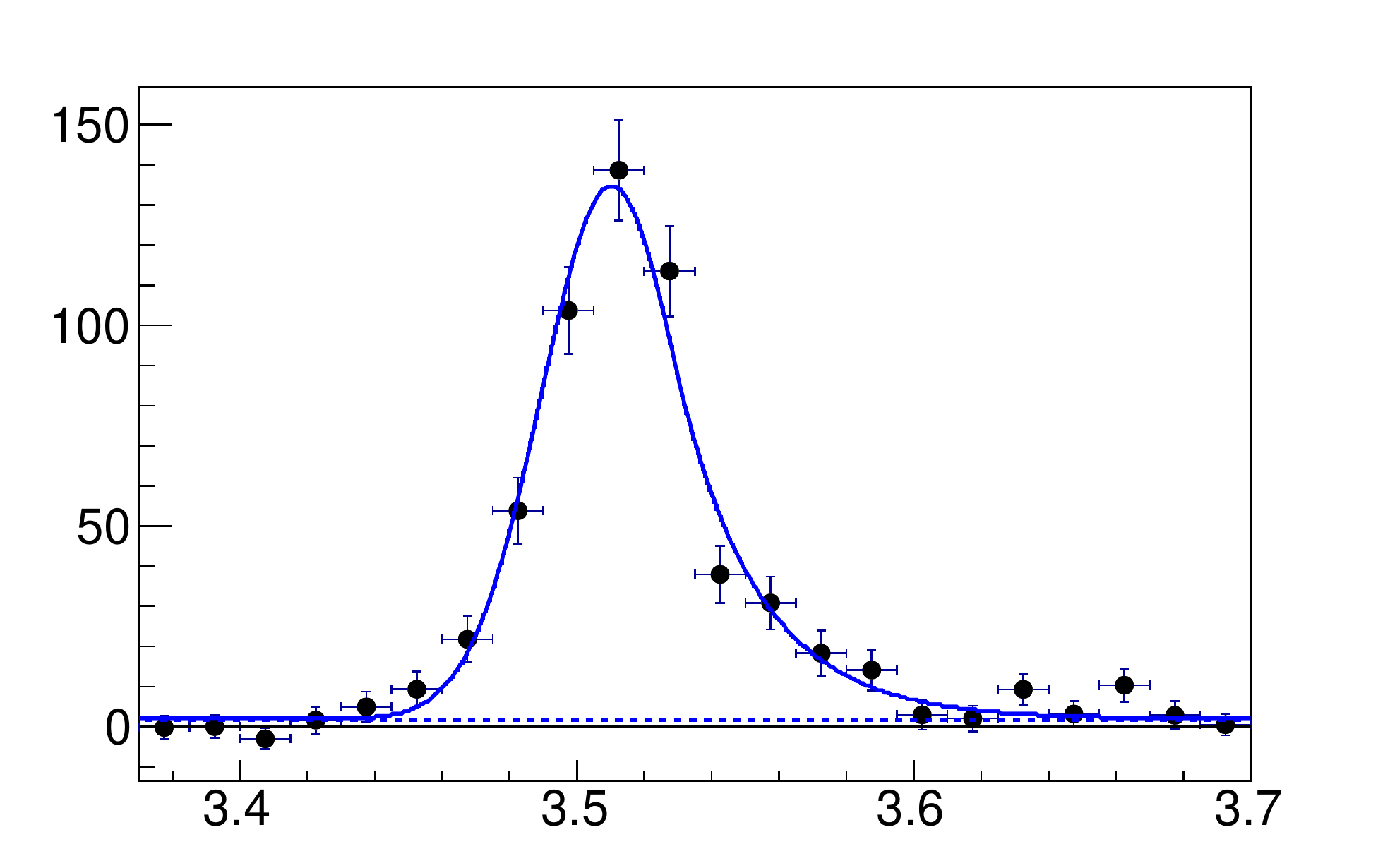}
    }

    \put(0,50){
      \includegraphics*[width=80mm,
      ]{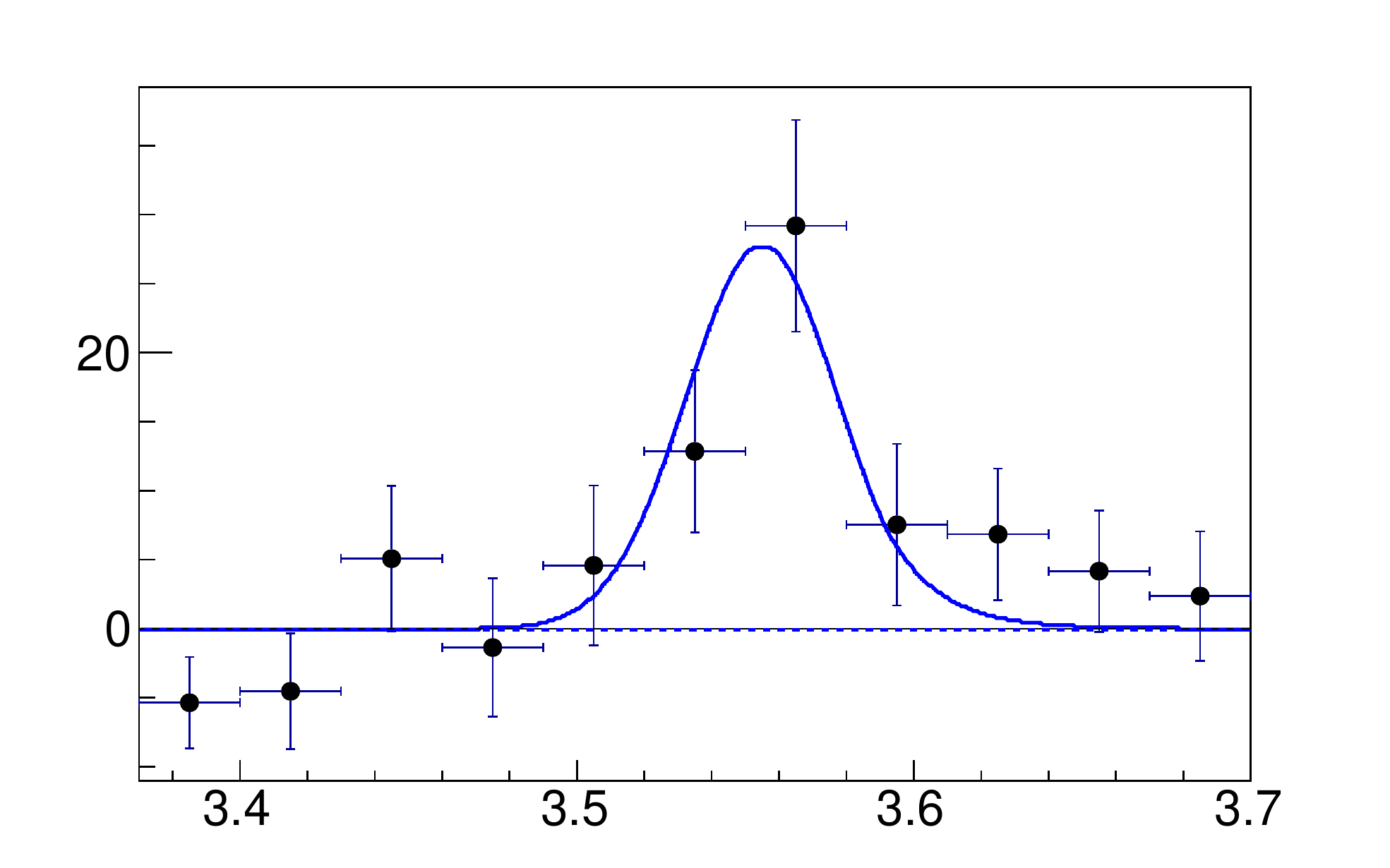}
    }

    \put(0,0){
      \includegraphics*[width=80mm,
      ]{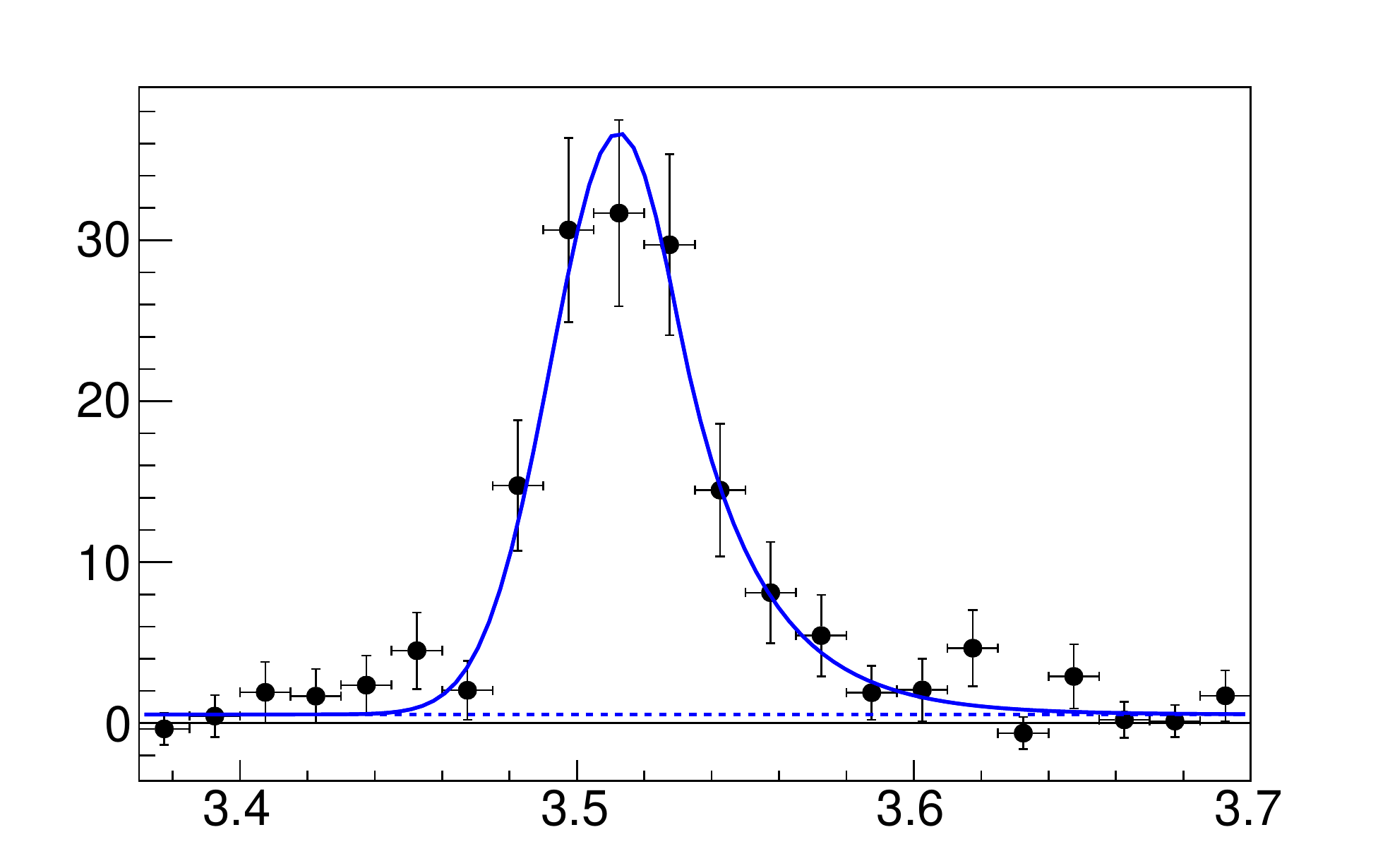}
    }

    \put(80,100){
      \includegraphics*[width=80mm,
      ]{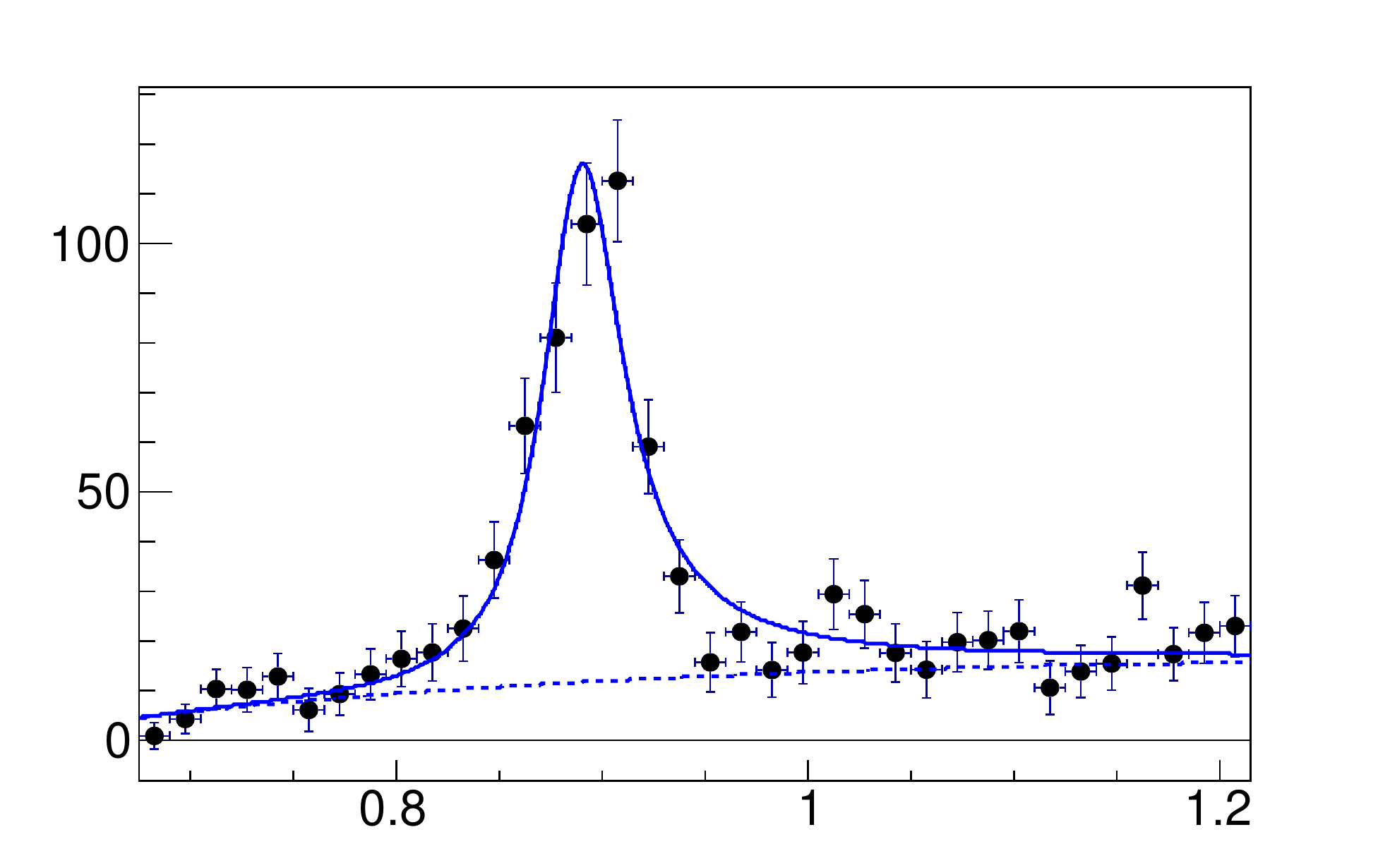}
    }

    \put(80,50){
      \includegraphics*[width=80mm,
      ]{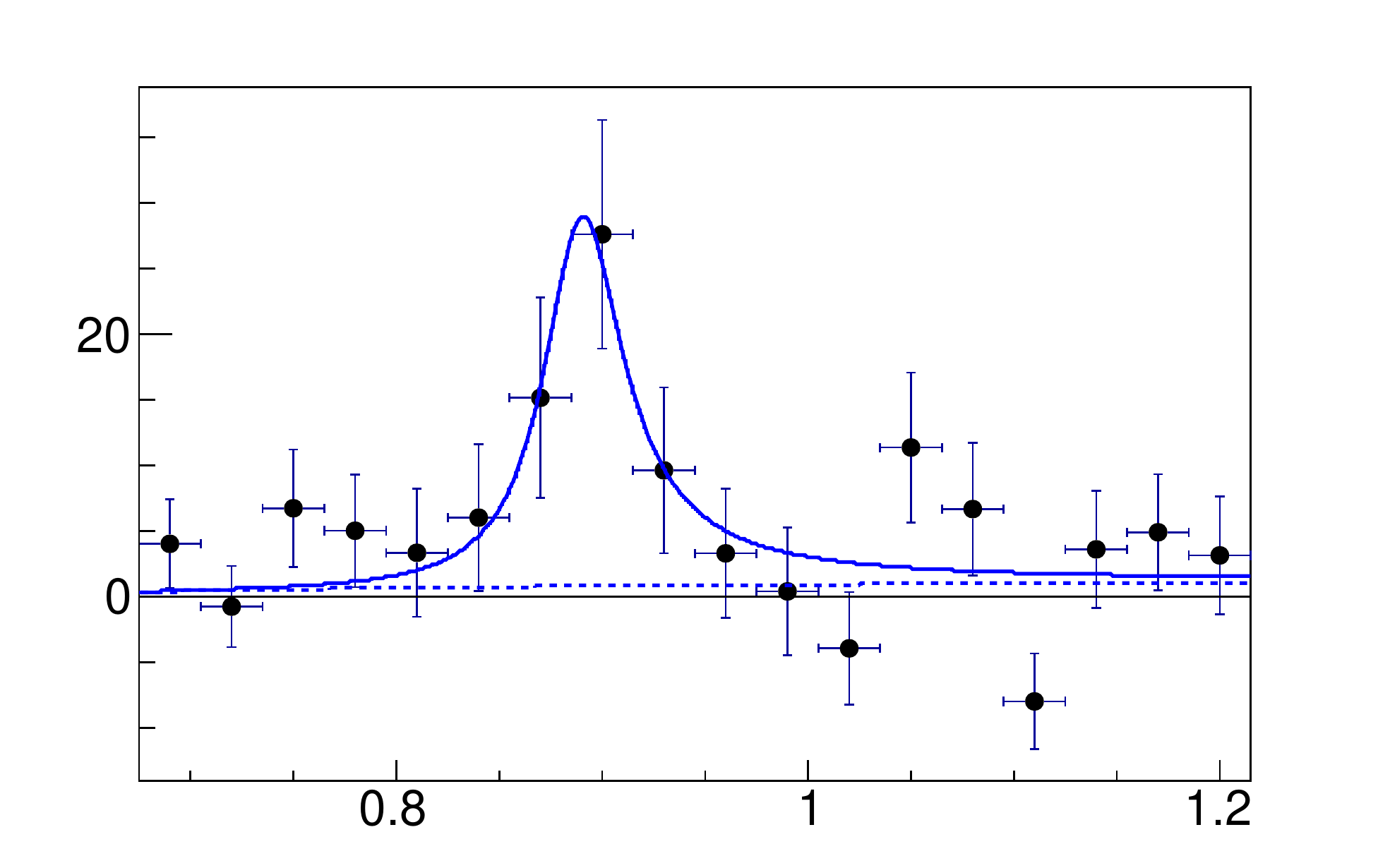}
    }

    \put(80,0){
      \includegraphics*[width=80mm,
      ]{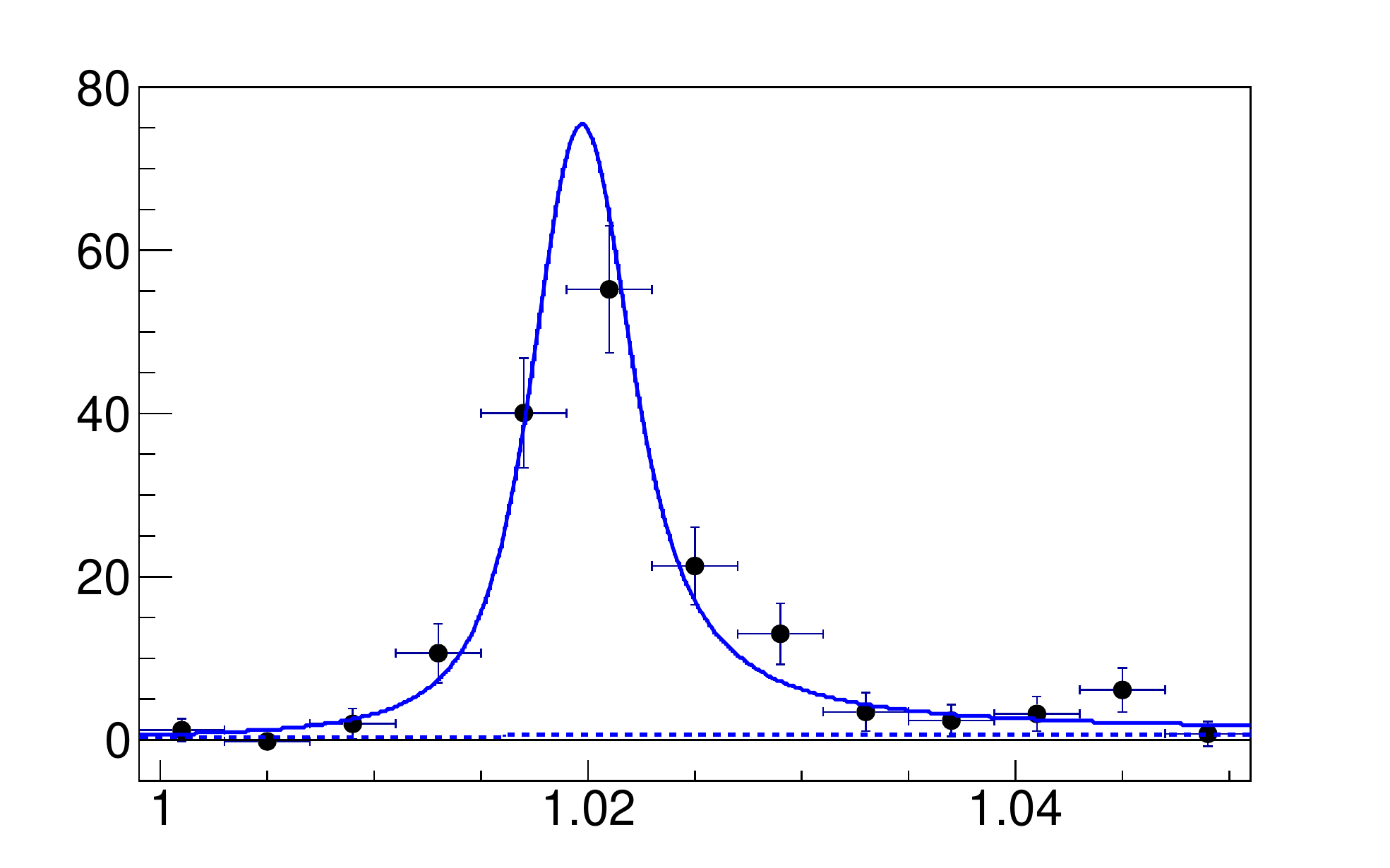}
    }

    \put(15,38){\small \lhcb}
    \put(95,38){\small \lhcb}
    \put(15,88){\small \lhcb}
    \put(95,88){\small \lhcb}
    \put(15,138){\small \lhcb}
    \put(95,138){\small \lhcb}

    \put(49,36){\small \footnotesize $\begin{array}{l} (\Pe) \\ \Bs\to\chicone\Pphi \end{array}$}
    \put(129,36){\small \footnotesize $\begin{array}{l} (\Pf) \\ \Bs\to\chicone\Pphi \end{array}$}
    \put(49,86){\small \footnotesize $\begin{array}{l} (\Pc) \\ \Bd\to\chictwo\Kstarz \end{array}$} 
    \put(129,86){\small \footnotesize $\begin{array}{l} (\Pd) \\ \Bd\to\chictwo\Kstarz \end{array}$}
    \put(49,136){\small \footnotesize $\begin{array}{l} (\Pa) \\ \Bd\to\chicone\Kstarz \end{array}$}
    \put(129,136){\small \footnotesize $\begin{array}{l} (\Pb) \\ \Bd\to\chicone\Kstarz \end{array}$}

    \put(38,-2){\scriptsize $\mathrm{M_{\jpsi\g}}$}
    \put(118,-2){\scriptsize $\mathrm{M_{\Kp\Km}}$}
    \put(38,48){\scriptsize $\mathrm{M_{\jpsi\g}}$}
    \put(118,48){\scriptsize $\mathrm{M_{\Kp\pim}}$}
    \put(38,98){\scriptsize $\mathrm{M_{\jpsi\g}}$}
    \put(118,98){\scriptsize $\mathrm{M_{\Kp\pim}}$}

    \put(62,-2){\scriptsize $\mathrm{[GeV/c^{2}]}$}
    \put(142,-2){\scriptsize $\mathrm{[GeV/c^{2}]}$}
    \put(62,48){\scriptsize $\mathrm{[GeV/c^{2}]}$}
    \put(142,48){\scriptsize $\mathrm{[GeV/c^{2}]}$}
    \put(62,98){\scriptsize $\mathrm{[GeV/c^{2}]}$}
    \put(142,98){\scriptsize $\mathrm{[GeV/c^{2}]}$}

    \begin{sideways}

    \put(10,80){\scriptsize Candidates / (15\mevcc)}
    \put(10,0){\scriptsize Candidates / (4\mevcc)}

    \put(60,80){\scriptsize Candidates / (30\mevcc)}
    \put(60,0){\scriptsize Candidates / (30\mevcc)}

    \put(110,80){\scriptsize Candidates / (15\mevcc)}
    \put(110,0){\scriptsize Candidates / (15\mevcc)}

    \end{sideways}

  \end{picture}

\caption{ \small Background-subtracted invariant mass distributions for: (a) $\jpsi\g$ and (b) $\Kp\pim$ final states from $\Bd\to\chicone\Kstarz$ 
decays obtained with the \chicone mass constraint applied to the $\Bd_{(\Ps)}$ candidate invariant mass; (c) $\jpsi\g$ and (d) $\Kp\pim$ final 
states from $\Bd\to\chictwo\Kstarz$ decays obtained with the \chictwo mass constraint applied to the $\Bd_{(\Ps)}$ candidate invariant mass; (e) 
$\jpsi\g$ and (f) $\Kp\Km$ final states from $\Bs\to\chicone\Pphi$ decays obtained with the \chicone mass constraint applied to the $\Bd_{(\Ps)}$ 
candidate invariant mass. The total fitted function (solid) and the non-resonant contribution (dotted) are shown. }

  \label{fig:sPlot_Bd2chi} \end{figure}

The resulting invariant mass distributions for $\jpsi\g$, $\Kp\pim$ and $\Kp\Km$ from $\Bd\to\Pchi_{\Pc}\Kstarz$ and $\Bs\to\chicone\Pphi$ candidates 
are shown in Fig.~\ref{fig:sPlot_Bd2chi}. The $\jpsi\g$ invariant mass distributions are modelled with the sum of a constant and a Crystal Ball 
function~\cite{CrystalBall1} with tail parameters fixed to simulation. In the \chictwo mode the signal peak position is fixed to the sum of the 
\chicone peak position and the known difference between \chicone and \chictwo masses~\cite{PDG}. The \chictwo mass resolution is fixed to the \chicone 
mass resolution multiplied by a scale factor determined using simulation. The $\Kp\pim$ and $\Kp\Km$ invariant mass distributions are modelled with 
the sum of a relativistic P-wave Breit-Wigner function with the natural width fixed to the known value~\cite{PDG} and a non-resonant component 
modelled with the LASS parametrization~\cite{LASS}. For the $\Kp\Km$ case the relativistic P-wave Breit-Wigner function is convolved with a Gaussian 
function 
for the detector resolution.


The signal peak positions are consistent with the known masses of the mesons while the invariant mass resolutions are consistent with the expectation 
from simulation. In the $\jpsi\g$ invariant mass distributions, the non-resonant contribution is consistent with zero. The resonant contributions for 
the $\Bd\to\chicone\Kstarz$ and $\Bs\to\chicone\Pphi$ decays are determined with the \chicone mass constraint while the resonant contribution for the 
$\Bd\to\chictwo\Kstarz$ decay is determined with the \chictwo mass constraint. The resulting resonant yields, obtained from the fits to the 
background-subtracted $\Kp\pim$ and $\Kp\Km$ distributions, are shown in Table~\ref{table:Yields}.

\begin{table}[b] \caption{ \small Signal yields for the $\B$ decays.} \begin{center} \begin{tabular}{lc}
        Decay & Yield \\
        \hline
        $\Bd\to\chicone\Kstarz$ & $\;\,\! \;\, 566 \pm 31$ \\
        $\Bd\to\chictwo\Kstarz$ & $\;\,\! \;\,\;\; 66 \pm 19$ \\
        $\Bs\to\chicone\Pphi$ & $\;\,\!  \;\, 146 \pm 14$ \\ 
        $\Bd\to\jpsi\Kstarz$ & $56,\!707 \pm 279 $ \\
        $\Bs\to\jpsi\Pphi$ & $15,\!027 \pm 139 $ \\

\end{tabular} \end{center} \label{table:Yields} \end{table}

\section[ $\text{\Bd\!\!\to\!\!\jpsi\!\!\Kstarz}$ and $\text{\Bs\!\!\to\!\!\jpsi\!\!\Pphi}$ decays ]
{ $\boldmath{\text{\Bd\!\!\to\!\!\jpsi\!\!\Kstarz}}$ and $\boldmath{\text{\Bs\!\!\to\!\!\jpsi\!\!\Pphi}}$ decays }

\label{sec:SigExBds_norm}

\begin{figure}[t]
  \setlength{\unitlength}{1mm}
  \centering
  \begin{picture}(160,40)

    \put(0,0){
      \includegraphics*[width=80mm,
      ]{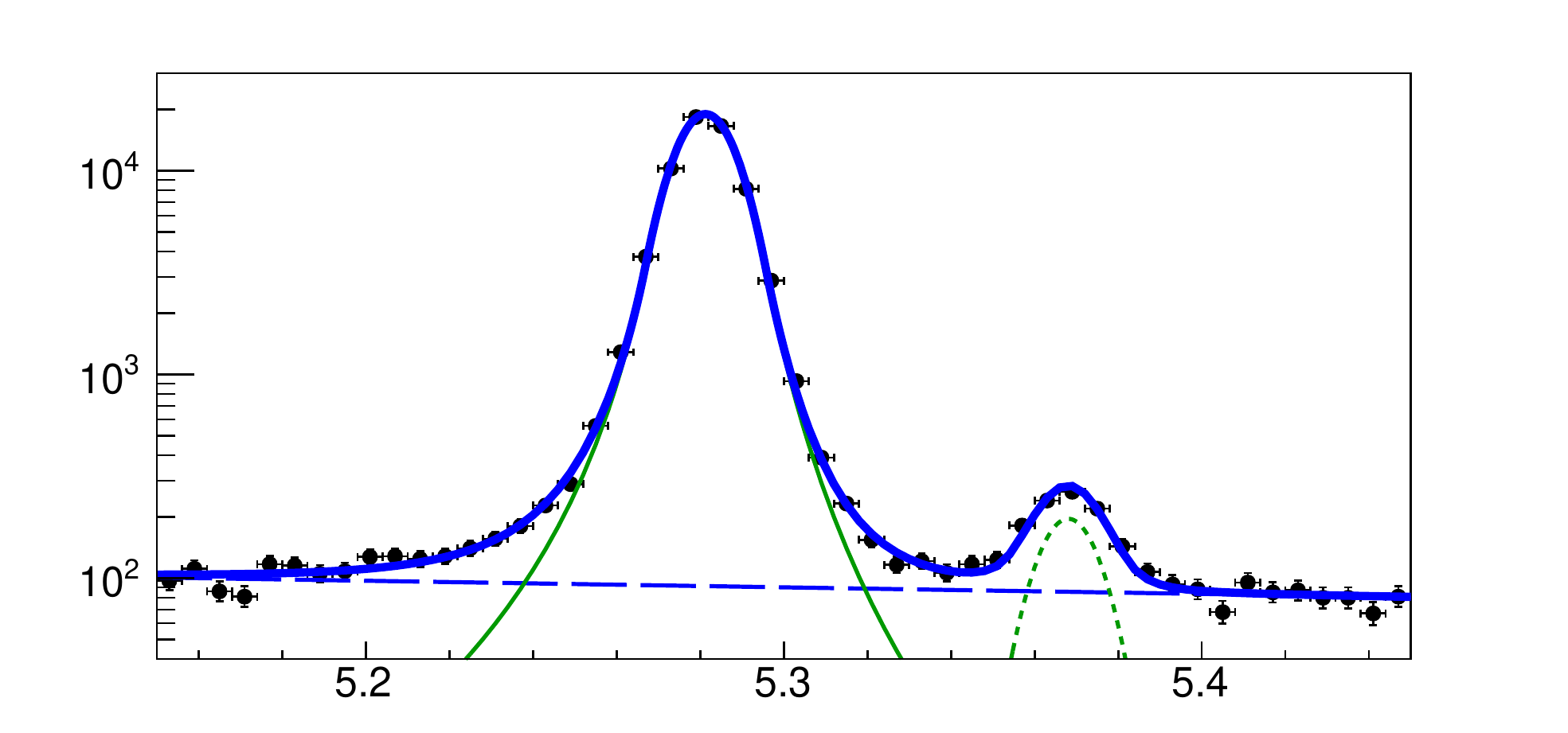}
    }
    \put(80,0){
      \includegraphics*[width=80mm,
      ]{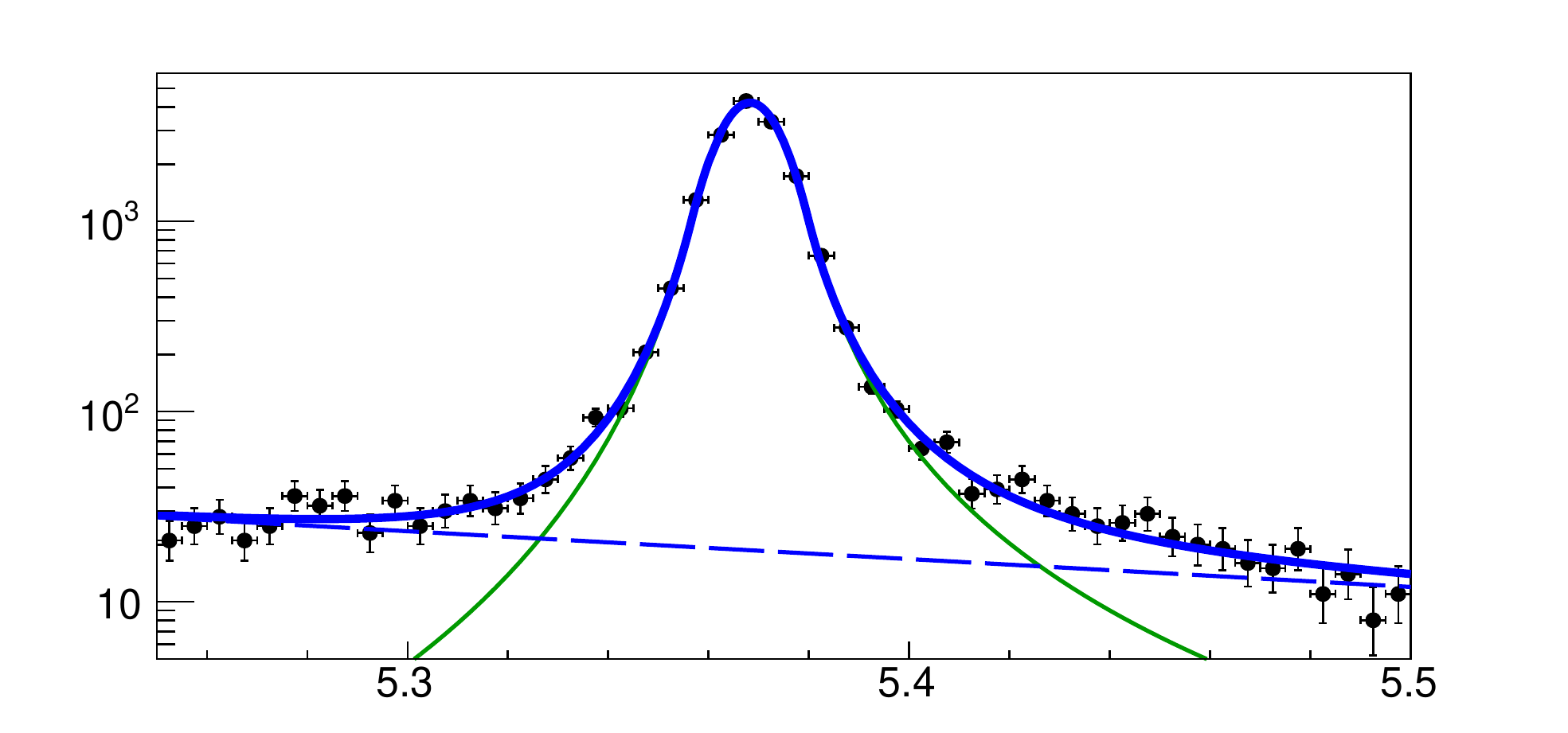}
    }

    \put(15,28){\small \lhcb}
    \put(95,28){\small \lhcb}

    \put(45,27){\small \footnotesize $\begin{array}{l} (\Pa) \\ \Bd\to\jpsi\Kstarz \end{array}$}
    \put(125,27){\small \footnotesize $\begin{array}{l} (\Pb) \\ \Bs\to\jpsi\Pphi \end{array}$}

    \put(38,-2){\scriptsize $\mathrm{M_{\jpsi\Kstarz}}$}
    \put(118,-2){\scriptsize $\mathrm{M_{\jpsi\Pphi}}$}

    \put(62,-2){\scriptsize $\mathrm{[GeV/c^{2}]}$}
    \put(142,-2){\scriptsize $\mathrm{[GeV/c^{2}]}$}

    \begin{sideways}

    \put(3,0){\scriptsize Candidates / (5\mevcc)}
    \put(3,80){\scriptsize Candidates / (5\mevcc)}
    \end{sideways}

  \end{picture}

  \caption { \small Invariant mass distributions for (a) $\Bd\to\jpsi\Kstarz$ and (b) $\Bs\to\jpsi\Pphi$. The total fitted function (thick solid 
blue), signal (thin solid green), the $\Bs\to\jpsi\Kstarz$ (green dotted) and the combinatorial background (dashed blue) are shown.}

  \label{fig:MassDistBds_norm}
\end{figure}

The $\Bd\to\chic\Kstarz$ and $\Bs\to\chicone\Pphi$ branching fractions are measured with respect to the $\Bd\to\jpsi\Kstarz$ and $\Bs\to\jpsi\Pphi$ 
decays to reduce the systematic uncertainties. The invariant mass distributions for the $\Bd\to\jpsi\Kstarz$ and $\Bs\to\jpsi\Pphi$ candidates after 
selection requirements are shown in Fig.~\ref{fig:MassDistBds_norm}. The signal and the $\Bs\to\jpsi\Kstarz$ invariant mass distributions are modelled 
by a double-sided Crystal Ball function and the combinatorial background is modelled by an exponential function. 
The parameters of the \Bs peak are fixed to be the same as those of the \Bd peak except the position and yield. 
The difference between the $\Bd\to\jpsi\Kstarz$ 
and $\Bs\to\jpsi\Kstarz$ peak positions is fixed to the world average~\cite{PDG}. 
The positions of the signal peaks are consistent with the known masses of the $\Bd_{(\Ps)}$ 
mesons~\cite{PDG} and the mass resolutions are consistent with expectations from simulation. 

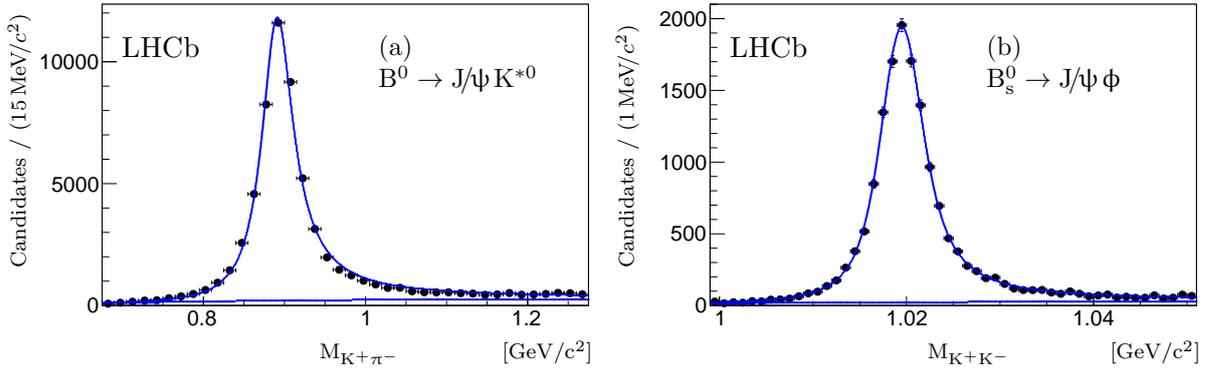
\begin{figure}[t]
  \setlength{\unitlength}{1mm}
  \centering
  \begin{picture}(160,50)

    \put(3,0){
      \includegraphics*[width=80mm
      ]{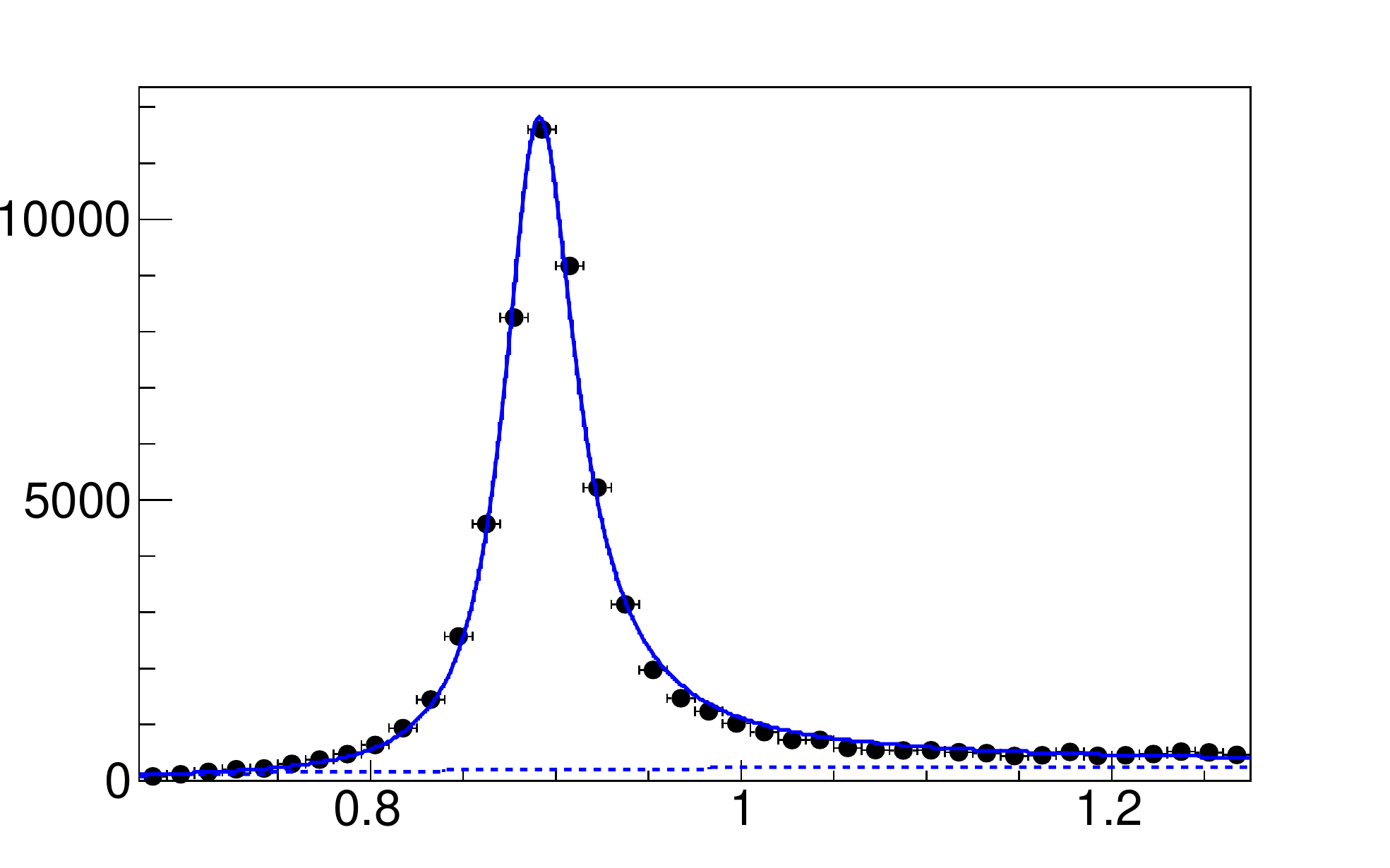}
    }
    \put(83,0){
      \includegraphics*[width=80mm
      ]{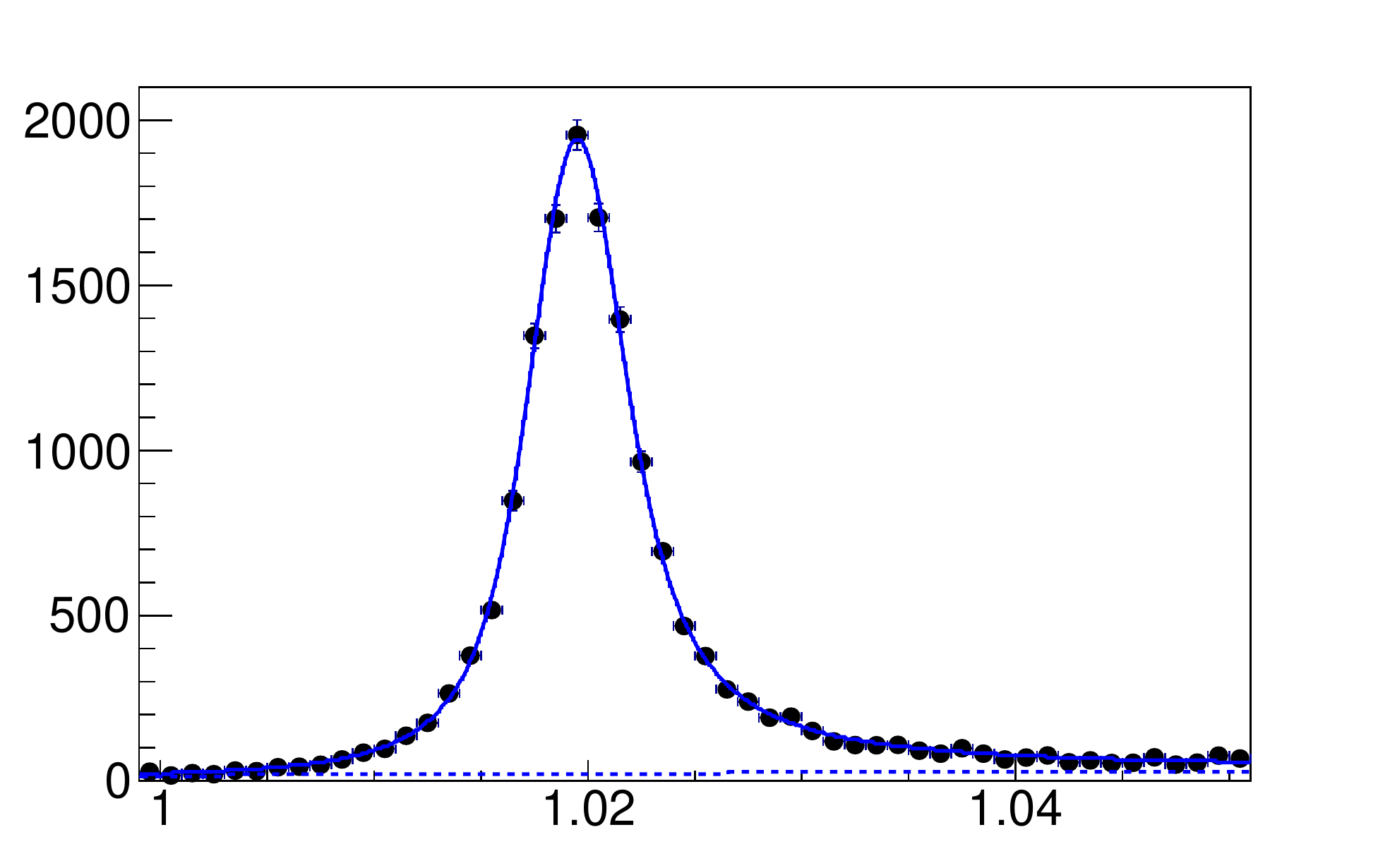}
    }

    \put(15,38){\small \lhcb}
    \put(95,38){\small \lhcb}

    \put(47,36){\small \footnotesize $\begin{array}{l} (\Pa) \\ \Bd\to\jpsi\Kstarz \end{array}$}
    \put(127,36){\small \footnotesize $\begin{array}{l} (\Pb) \\ \Bs\to\jpsi\Pphi \end{array}$}
  
    \put(41,-2){\scriptsize $\mathrm{M_{\Kp\pim}}$}
    \put(121,-2){\scriptsize $\mathrm{M_{\Kp\Km}}$}

    \put(65,-2){\scriptsize $\mathrm{[GeV/c^{2}]}$}
    \put(145,-2){\scriptsize $\mathrm{[GeV/c^{2}]}$}

    \begin{sideways}

    \put(10,0){\scriptsize Candidates / (1\mevcc)}
    \put(10,80){\scriptsize Candidates / (15\mevcc)}

    \end{sideways}

  \end{picture}
  \caption { \small Background-subtracted invariant mass distributions for (a) $\Kp\pim$ combinations from $\Bd\to\jpsi\Kstarz$ decays and (b) 
$\Kp\Km$ combination from $\Bs\to\jpsi\Pphi$ decays. The total fitted function (solid) and the non-resonant contribution (dotted) are shown.}

  \label{fig:sPlot_Bds_norm}
\end{figure}

The resonant contributions in the $\Bd\to\jpsi\Kstarz$ and $\Bs\to\jpsi\Pphi$ decays are determined using the {\em sPlot} technique with the same 
method as that used for the $\Bd\to\Pchi_{\Pc}\Kstarz$ and $\Bs\to\Pchi_{\Pc}\Pphi$ decays. The resulting $\Kp\pim$ and $\Kp\Km$ invariant mass 
distributions from 
$\Bd\to\jpsi\Kstarz$ and $\Bs\to\jpsi\Pphi$ candidates are shown in Fig.~\ref{fig:sPlot_Bds_norm}. The resulting resonant yields are summarized in 
Table~\ref{table:Yields}. 
The S-wave contributions are consistent with those considered in other analyses~\cite{B2psiX_paper,SwaveKpi,SwaveKK}.

\section{Efficiencies and systematic uncertainties}
\label{sec:Measurement}

The branching fraction ratios are calculated using the formulas

\begin{equation}
\displaystyle{
\begin{array}{lll}

\frac{\displaystyle \BR(\B\to\chicone\PX)}{\displaystyle \BR(\B\to\jpsi\PX)} &=& \frac{\displaystyle N_{\B\to\chicone\PX}}{\displaystyle 
N_{\B\to\jpsi\PX}} \times \frac{\displaystyle \Pvarepsilon_{\B\to\jpsi\PX}}{\displaystyle \Pvarepsilon_{\B\to\chicone\PX}} \times 
\frac{\displaystyle 1}{\displaystyle \BR(\chicone\to\jpsi\g)} ~, \\[1.3em]

\frac{\displaystyle \BR(\B\to\chictwo\PX)}{\displaystyle \BR(\B\to\chicone\PX)} &=& \frac{\displaystyle N_{\B\to\chictwo\PX}}{\displaystyle 
N_{\B\to\chicone\PX}} \times \frac{\displaystyle \Pvarepsilon_{\B\to\chicone\PX}}{\displaystyle \Pvarepsilon_{\B\to\chictwo\PX}} \times 
\frac{\displaystyle \BR(\chicone\to\jpsi\g)}{\displaystyle \BR(\chictwo\to\jpsi\g)} ~,

\end{array}
}
\label{eq:Ratio1}
\end{equation}

\noindent where $N$ represents the measured yield and $\Pvarepsilon$ represents the total efficiency. The total efficiency is the product of the 
geometrical acceptance, the detection, reconstruction, selection and trigger efficiencies. The efficiencies are 
derived using simulation and are presented in Table~\ref{table:AllEff}.

\begin{table}[b]
\caption{\small Total efficiencies for all decay modes. Uncertainties are statistical only and reflect the size of the simulation sample.}
\begin{center}
\begin{tabular}{lc}


        Decay & Efficiency~$[10^{-4}]$ \\
        \hline
        $\Bd\to\chicone\Kstarz$ & $\;\, 7.89 \pm 0.12 $ \\
        $\Bd\to\chictwo\Kstarz$ & $\;\, 9.45 \pm 0.13 $ \\
        $\Bs\to\chicone\Pphi$   & $12.7 \;\, \pm 0.2 \;\,$ \\
        $\Bd\to\jpsi\Kstarz$    & $53.9 \;\, \pm 0.3 \;\,$ \\
        $\Bs\to\jpsi\Pphi$      & $85.1 \;\, \pm 0.4 \;\,$ \\

\end{tabular}
\end{center} 
\label{table:AllEff}
\end{table}




Most potential sources of systematic uncertainty cancel in the ratio, in particular, those related to the muon and \jpsi reconstruction and 
identification. The remaining systematic uncertainties are summarized in Table~\ref{table:SysUnc} and each is now discussed in turn.


Systematic uncertainties related to the signal determination procedure are estimated using a number
of alternative options. For each of the alternatives the ratio of event yields is calculated and
the systematic uncertainty is then determined as the maximum deviation of this ratio from the ratio obtained with the baseline model. 
For the $\Bd_{(\Ps)}$ meson decays a fit with a 
second-order polynomial for the combinatorial background description, a fit with a Crystal Ball~\cite{CrystalBall1} function for the 
signal peaks and fit over different ranges of invariant mass are used. In the \Bs channel a fit including the \chictwo decay mode is also performed. 
For the $\Kp\pim$ and $\Kp\Km$ combinations the fits are repeated, modelling the background with an S-wave two-body phase-space function or an S-wave 
two-body phase-space function multiplied by a linear function. The $\Kp\pim$ and $\Kp\Km$ invariant mass ranges and the bin size are also varied. 
The resulting uncertainties are 3\% on ${\BR(\Bd\to\chicone\Kstarz)/\BR(\Bd\to\jpsi\Kstarz)}$, $5\%$ on 
${\BR(\Bs\to\chicone\Pphi)/\BR(\Bs\to\jpsi\Pphi)}$, 
and $9\%$ on ${\BR(\Bd\to\chictwo\Kstarz)/\BR(\Bd\to\chicone\Kstarz)}$. 

Another important source of systematic uncertainty arises from the potential disagreement between  data and simulation in the estimation of 
efficiencies. To study this source of uncertainty, the selection criteria are varied in ranges corresponding to as much as $30\%$ change in the signal 
yields and the ratios of the selection and reconstruction efficiencies are compared between data and simulation. The largest difference ($3\%$) is 
assigned as a systematic uncertainty in each mode. 

A further source of possible disagreement between data and simulation is the photon reconstruction efficiency.
As in Ref.~\cite{Dasha_paper}, the photon reconstruction efficiency has been studied using $\Bp\to\jpsi\Kstarp$, followed by $\Kstarp\to\Kp\piz$ and 
$\piz\to\g\g$ decays. For photons with transverse momentum greater than 0.7\gevc the agreement between data and 
simulation is at the level of 4\%, which is assigned as a systematic uncertainty to the ratios
${\BR(\Bd\to\chicone\Kstarz)/\BR(\Bd\to\jpsi\Kstarz)}$ and ${\BR(\Bs\to\chicone\Pphi)/\BR(\Bs\to\jpsi\Pphi)}$. As the transverse momentum 
spectra of photons are similar in $\Bd\to\chicone\Kstarz$ and $\Bd\to\chictwo\Kstarz$ decays, this systematic uncertainty cancels in the ratio  
${\BR(\Bd\to\chictwo\Kstarz)/\BR(\Bd\to\chicone\Kstarz)}$.

The systematic uncertainty related to the trigger efficiency has been obtained by comparing the trigger efficiency ratios in data and simulation for 
the high yield decay modes $\Bp\to\jpsi\Kp$ and $\Bp\to\Ppsi(2\PS)\Kp$ which have similar kinematics and the same trigger requirements as the channels 
under study in this analysis~\cite{B2psiX_paper}. An agreement within 1\% is found, which is assigned as systematic uncertainty. 

The uncertainty due to the finite simulation sample size is included in the statistical uncertainty of the result by adding it in quadrature to the
statistical uncertainty on the ratio of yields. 

\begin{table}[htb]
\caption{ \small Relative systematic uncertainties (in $\%$) on the ratio of branching fractions.}
\begin{center}
\begin{tabular}{lccc}

	Source 		
& $\frac{\BR(\Bs\to\chicone\Pphi)}{\BR(\Bs\to\jpsi\Pphi)}$ 
& $\frac{\BR(\Bd\to\chicone\Kstarz)}{\BR(\Bd\to\jpsi\Kstarz)}$ 
& $\frac{\BR(\Bd\to\chictwo\Kstarz)}{\BR(\Bd\to\chicone\Kstarz)}$ 
\\
	\hline

	Signal determination			& $5$	& $3$	& $9$ \\ 

	Efficiencies from simulation		& $3$	& $3$	& $3$ \\ %
	Photon reconstruction			& $4$	& $4$	& $-$ \\ 

	Trigger					& $1$	& $1$	& $1$ \\ %

	\hline
	Sum in quadrature			& $7$	& $6$	& $10$ 

\end{tabular}
\end{center}
\label{table:SysUnc}
\end{table}

\section{ Results and summary}
\label{sec:Result}

The first observation of the $\Bs\to\chicone\Pphi$ decay has been made with a data sample, corresponding to an integrated luminosity of 1.0\invfb of 
$\proton\proton$ collisions at a centre-of-mass energy of 7\tev, collected with the LHCb detector. Its branching fraction, normalized to that 
of the $\Bs\to\jpsi\Pphi$ decay and using the known value $\BR(\chicone\to\jpsi\g) = (34.4 \pm 1.5)\%$~\cite{PDG}, is measured to be

$$
\begin{array}{llll}

\dfrac{ \BR(\Bs\to\Pchi_{\Pc1}\Pphi)}{ \BR(\Bs\to\jpsi\Pphi )} &=& (6.51~\pm0.64\,\stat\pm0.46\,\syst) \times 10^{-2} \times 
\dfrac{1}{\BR(\chicone\to\jpsi\g)} &= \\
\noalign{\vskip 3pt}

&=& (18.9~\pm1.8\,\stat\pm1.3\,\syst\pm0.8\,(\BR)) \times 10^{-2},

\end{array}
$$

\noindent where the third uncertainty corresponds to the uncertainty on the branching 
fraction of the $\chicone\to\jpsi\g$ decay. Using the same 
dataset, the ratio of the branching fractions of the $\Bd\to\chicone\Kstarz$ and $\Bd\to\jpsi\Kstarz$ modes and the ratio of the 
branching fractions of the $\Bd\to\chictwo\Kstarz$ and $\Bd\to\chicone\Kstarz$ modes have been measured. The ratios are determined using 
Eq.~\ref{eq:Ratio1} and the known value $\frac{\BR(\Pchi_{\Pc1}\to\jpsi\g)}{\BR(\Pchi_{\Pc2}\to\jpsi\g)} = \frac{(34.4\pm1.5)\%}{(19.5\pm0.8)\%} =
1.76\pm0.11$~\cite{PDG} and are

$$\begin{array}{llll}

\dfrac{ \BR(\Bd\to\Pchi_{\Pc1}\Kstarz)}{ \BR(\Bd\to\jpsi\Kstarz )} &=& (6.82~\pm0.39\,\stat\pm0.41\,\syst) \times 10^{-2} \times
\dfrac{1}{\BR(\chicone\to\jpsi\g)} &= \\
\noalign{\vskip 3pt}
 &=& (19.8~\pm1.1\,\stat\pm1.2\,\syst\pm0.9\,(\BR)) \times 10^{-2} , \\
\noalign{\vskip 5pt}

\dfrac{ \BR(\Bd\to\Pchi_{\Pc2}\Kstarz)}{ \BR(\Bd\to\Pchi_{\Pc1}\Kstarz )} &=& (9.74~\pm2.86\,\stat\pm0.97\,\syst) \times 10^{-2} \times
\dfrac{\BR(\chicone\to\jpsi\g)}{\BR(\chictwo\to\jpsi\g)} &=\\
\noalign{\vskip 3pt}
 &=& (17.1~\pm5.0\,\stat\pm1.7\,\syst\pm1.1\,(\BR)) \times 10^{-2},

\end{array}
$$

\noindent  where the third uncertainty is due to the uncertainty on the branching fractions of the ${\Pchi_{\Pc}\to\jpsi\g}$ modes. 

The ratio $\BR(\Bd\to\chicone\Kstarz)/\BR(\Bd\to\jpsi\Kstarz)$ obtained in this paper is compatible with, but more precise than,
the previous best value of $(17.2^{+3.6}_{-3.0}) {\times 10^{-2}}$ determined from the world average value $\BR(\Bd\to\chicone\Kstarz) = 
(2.22^{+0.40}_{-0.31}) \times 10^{-4}$~\cite{PDG} and the branching fraction $\BR(\Bd\to\jpsi\Kstarz) = (1.29\pm 0.05\pm 0.13) \times 10^{-3}$ 
measured by the \belle collaboration~\cite{belle_jpsi}. 
Other measurements of $\BR(\Bd\to\jpsi\Kstarz)$ are not considered as they do not take into account the $\Kp\pim$ S-wave component. 
The ratio ${\BR(\Bd\to\Pchi_{\Pc2}\Kstarz)/\BR(\Bd\to\Pchi_{\Pc1}\Kstarz)}$ obtained in this paper is compatible with the value derived 
from \babar measurements, $(26 \pm 7\stat) \times 10^{-2}$~\cite{babar},  taking only the statistical uncertainties into account.


\section*{Acknowledgements}

\noindent We express our gratitude to our colleagues in the CERN
accelerator departments for the excellent performance of the LHC. We
thank the technical and administrative staff at the LHCb
institutes. We acknowledge support from CERN and from the national
agencies: CAPES, CNPq, FAPERJ and FINEP (Brazil); NSFC (China);
CNRS/IN2P3 and Region Auvergne (France); BMBF, DFG, HGF and MPG
(Germany); SFI (Ireland); INFN (Italy); FOM and NWO (The Netherlands);
SCSR (Poland); ANCS/IFA (Romania); MinES, Rosatom, RFBR and NRC
``Kurchatov Institute'' (Russia); MinECo, XuntaGal and GENCAT (Spain);
SNSF and SER (Switzerland); NAS Ukraine (Ukraine); STFC (United
Kingdom); NSF (USA). We also acknowledge the support received from the
ERC under FP7. The Tier1 computing centres are supported by IN2P3
(France), KIT and BMBF (Germany), INFN (Italy), NWO and SURF (The
Netherlands), PIC (Spain), GridPP (United Kingdom). We are thankful
for the computing resources put at our disposal by Yandex LLC
(Russia), as well as to the communities behind the multiple open
source software packages that we depend on.

\addcontentsline{toc}{section}{References}
\bibliographystyle{LHCb}
\bibliography{main}

\end{document}